\documentclass[sigconf,10pt,screen]{acmart}
\AtBeginDocument{%
  }

\copyrightyear{2025}
\acmYear{2025}
\setcopyright{cc}
\setcctype{by}
\acmConference[ICS '25]{2025 International Conference on Supercomputing}{June 8--11, 2025}{Salt Lake City, UT, USA}
\acmBooktitle{2025 International Conference on Supercomputing (ICS '25), June 8--11, 2025, Salt Lake City, UT, USA}
\acmDOI{10.1145/3721145.3725744}
\acmISBN{979-8-4007-1537-2/2025/06}

\usepackage{acmart-taps}

\def\conference

\usepackage{etoolbox}
\newcommand{\ifconference}[1]{{{\ifx\fullversion\undefined{#1}\fi}\xspace}}
\newcommand{\iffullversion}[1]{{{\ifx\conference\undefined{#1}\fi}\xspace}}

\usepackage{graphicx}  
\usepackage{lipsum}  
\newcommand{\hide}[1]{} 
\usepackage{xspace}
\usepackage{textcomp}
\usepackage{comment} 
\usepackage{verbatim}

\ifx\nocomment\undefined
\newcommand{\yan}[1]{{\textcolor{violet}{Yan: #1}}}
\newcommand{\yihan}[1]{{\color{blue}{\bf Yihan:} #1}}
\newcommand{\xiaojun}[1]{{\color{cyan}{\bf Xiaojun:} #1}}
\newcommand{\zijin}[1]{{\textcolor{purple}{Zijin: #1}}}
\newcommand{\letong}[1]{{\textcolor{orange}{Letong: #1}}}

\else
\newcommand{\yan}[1]{}
\newcommand{\yihan}[1]{}
\newcommand{\xiaojun}[1]{}
\newcommand{\zijin}[1]{}
\newcommand{\letong}[1]{}

\fi

\usepackage{url}

\newcommand{\defn}[1]{\emph{\textbf{#1}}} 
\newcommand{\emp}[1]{\emph{\boldmath\textbf{#1}\unboldmath}} 
\newcommand{\fname}[1]{\textsc{#1}} 
\newcommand{\vname}[1]{\mathit{#1}} 

\usepackage[ruled,lined,linesnumbered,noend]{algorithm2e}
\usepackage[noend]{algpseudocode}

\makeatletter
\patchcmd{\@algocf@start}
  {-1.5em}
  {0pt}
  {}{}
\newcommand{\nosemic}{\renewcommand{\@endalgocfline}{\relax}}
\newcommand{\dosemic}{\renewcommand{\@endalgocfline}{\algocf@endline}}

\SetSideCommentLeft
\SetKwInput{notations}{Notations}
\SetKwInput{notes}{Notes}
\SetKwInput{maintains}{Maintains}

\SetKwProg{myfunc}{Function}{}{}
\SetKwFor{parForEach}{ParallelForEach}{do}{endfor}
\SetKwFor{Justrepeat}{Repeat}{}{}

\SetKw{MIN}{min}
\SetKw{MAX}{max}
\SetKw{OR}{or}
\SetKw{AND}{and}

\SetCommentSty{mycommfont}

\usepackage{cleveref}
\crefname{section}{Sec.}{Sec.}
\crefname{theorem}{Thm.}{Thm.}
\crefname{lemma}{Lem.}{Lem.}
\crefname{corollary}{Col.}{Col.}
\crefname{table}{Tab.}{Tab.}
\crefname{algorithm}{Alg.}{Alg.}
\crefname{figure}{Fig.}{Fig.}
\crefname{fact}{Fact}{Fact}
\Crefname{table}{Tab.}{Tab.}
\crefname{problem}{Problem}{Problem}

\let \originalleft \left
\let\originalright\right
\renewcommand{\left}{\mathopen{}\mathclose\bgroup\originalleft}
\renewcommand{\right}{\aftergroup\egroup\originalright}

\usepackage{scalerel} 

\newtheoremstyle{exampstyle}
{.5em} 
{1em} 
{\it} 
{.5em} 
{\it \bfseries} 
{.} 
{.5em} 
{} 
\theoremstyle{exampstyle} 
\theoremstyle{exampstyle} 
\theoremstyle{exampstyle} 
\theoremstyle{exampstyle} 

\makeatletter

\newcommand{\R}{\mathbb{R}}

\newcommand{\whp}[1]{\emph{whp}}

\DeclareMathOperator*{\argmin}{\arg\!\min}

\newcommand{\true}{\emph{true}}
\newcommand{\false}{\emph{false}}

\usepackage{pifont}

\newcommand{\modelop}[1]{\texttt{#1}}
\newcommand{\forkins}{\modelop{fork}}

\newcommand{\thread}{thread}

\newcommand{\insformat}[1]{\texttt{#1}}

\newcommand{\cas}{{\insformat{compare\_and\_swap}}}

\newcommand{\WriteMin}{\insformat{write\_min}\xspace}

\usepackage[shortlabels]{enumitem}

\setlist{topsep=0.3em,itemsep=0.2em,parsep=0.1em,leftmargin=*}

\usepackage{float}
\usepackage[labelfont=bf,font={small},aboveskip=0em, belowskip=0em]{caption}

\usepackage[labelfont=bf,list=true,skip=0em]{subcaption}
\usepackage[rightcaption]{sidecap}

\usepackage{wrapfig}

\usepackage{array}
\newcolumntype{L}[1]{>{\raggedright\let\newline\\\arraybackslash\hspace{0pt}}m{#1}}
\newcolumntype{C}[1]{>{\centering\let\newline\\\arraybackslash\hspace{0pt}}m{#1}}
\newcolumntype{R}[1]{>{\raggedleft\let\newline\\\arraybackslash\hspace{0pt}}m{#1}}
\newcolumntype{B}{>{\bf}c}
\usepackage{rotating}

\usepackage{booktabs} 
\usepackage{multicol,multirow}
\usepackage{longtable} 
\usepackage{supertabular} 
\usepackage{colortbl}
\usepackage{bigstrut}

\newcommand{\myparagraph}[1]{\vspace{.1em}\noindent\emp{#1}\enspace}

\usepackage{framed}

\usepackage{listings}

\newdimen\zzsize
\newdimen\kwsize
\newcommand{\basicstyle}{\fontsize{\zzsize}{1\zzsize}\ttfamily}
\newcommand{\keywordstyle}{\fontsize{\kwsize}{1\kwsize}\ttfamily\bf}

\newdimen\zzlstwidth
\usepackage{tikz} 

  \newcommand{\oursys}{\textsf{SPoCH}}
  \newcommand{\oursysTab}{\textsf{Ours}}

  \newcommand{\step}[1]{\textsf{#1}}
  \newcommand{\degree}{\mathit{d}\xspace}
  
  \newcommand{\priority}{\mathit{P}}

  \newcommand{\overlayG}{{\mathit{G_{\scriptsize \mbox{\it O}}}}}
  \newcommand{\overlayE}{{\mathit{E_{\scriptsize \mbox{\it O}}}}}
  \newcommand{\overlayV}{\mathit{V_{\scriptsize \mbox{\it O}}}}
  \newcommand{\residualE}{\mathit{E_{\scriptsize \mbox{\it CH}}}}
  
  \newcommand{\CH}{\mathit{G_{\scriptsize \mbox{\it CH}}}}
  \newcommand{\PCH}{\mathit{G_{\scriptsize \mbox{\it CH}}}}
  \newcommand{\ForwardEdges}{\mathit{\residualE^{\uparrow}}}
  \newcommand{\BackwardEdges}{\mathit{\residualE^{\downarrow}}}
  \newcommand{\ShortcutEdges}{\mathit{E^{+}}}
  \newcommand{\ECH}{\mathit{\residualE}}
  \newcommand{\FeasibleCandidates}{\mathit{V_F}}
  \newcommand{\AffectedNeighbors}{\mathit{V_A}}
  \newcommand{\PruneVertices}{\mathit{V_W}}
  \newcommand{\WPSSources}{\mathit{V_W}}
  \newcommand{\WPSSourcesinit}{\mathit{V_S}}
  \newcommand{\dist}{\vname{dist}}
  
  \newcommand{\selectFraction}{\mathit{\beta}}
  \newcommand{\batchInsert}{\mathit{insert}}
  \newcommand{\nin}{N_{\scriptsize \mbox{\it in}}}
  \newcommand{\nout}{N_{\scriptsize \mbox{\it out}}}
  \newcommand{\n}{N}

  \newcommand{\prune}{\step{LSM}}

  \newcommand{\rank}{\mathit{\pi}}

  \newcommand{\impname}[1]{\textsf{#1}}
  \newcommand{\rk}{\impname{RK}}
  \newcommand{\RK}{\rk}
  \newcommand{\routingkit}{\impname{RoutingKit}}
  \newcommand{\RoutingKit}{\routingkit}
  \newcommand{\chc}{\impname{CC}}
  \newcommand{\CC}{\chc}
  \newcommand{\chconstructor}{\impname{CH-Constructor}}
  \newcommand{\OSRM}{\impname{OSRM}}
  \newcommand{\osrm}{\OSRM}
  \newcommand{\PHAST}{\impname{PHAST}}
  \newcommand{\CHASE}{\impname{CHASE}}
  \newcommand{\phast}{\PHAST}

\begin{document}

\title[Parallel Contraction Hierarchies Can Be Efficient and Scalable]
{Parallel Contraction Hierarchies Can Be \\ Efficient and Scalable}

\author{Zijin Wan}
\affiliation{%
  \institution{University of California, Riverside}
  \city{Riverside}
  \state{CA}
  \country{USA}
}
\email{zijin.wan@email.ucr.edu}

\author{Xiaojun Dong}
\affiliation{%
  \institution{University of California, Riverside}
  \city{Riverside}
  \state{CA}
  \country{USA}
}
\email{xdong038@ucr.edu}

\author{Letong Wang}
\affiliation{%
  \institution{University of California, Riverside}
  \city{Riverside}
  \state{CA}
  \country{USA}
}
\email{lwang323@ucr.edu}

\author{Enzuo Zhu}
\affiliation{%
  \institution{University of California, Davis}
  \city{Davis}
  \state{CA}
  \country{USA}
}
\email{ezzhu@ucdavis.edu}

\author{Yan Gu}
\affiliation{%
  \institution{University of California, Riverside}
  \city{Riverside}
  \state{CA}
  \country{USA}
}
\email{ygu@cs.ucr.edu}

\author{Yihan Sun}
\affiliation{%
  \institution{University of California, Riverside}
  \city{Riverside}
  \state{CA}
  \country{USA}
}
\email{yihans@cs.ucr.edu}

\begin{abstract}
Contraction Hierarchies (CH) (Geisberger et al., 2008) is one of the most widely used algorithms
for shortest-path queries on road networks.
Compared to Dijkstra's algorithm, CH enables orders of magnitude
faster query performance through a preprocessing phase,
which iteratively categorizes vertices into hierarchies and adds shortcuts.
However, constructing a CH is an expensive task.
Existing solutions, including parallel ones, may suffer from long construction time.
Especially, in our experiments, we observe that existing parallel solutions demonstrate unsatisfactory scalability,
and have performance close to sequential algorithms.

We present \oursys{} (\textbf{S}calable \textbf{P}arallelization \textbf{o}f \textbf{C}ontraction \textbf{H}ierarchies),
an efficient and scalable CH construction algorithm in parallel.
To address the challenges in previous work, our improvements focus on both redesigning the algorithm
and leveraging parallel data structures. 
We compare \oursys{} with the state-of-the-art sequential and parallel implementations
on 16 graphs of various types.
Our experiments show that \oursys{} achieves $11$--$68\times$ speedups over the best sequential baseline and $3.8$--$41\times$ speedups over the best parallel baseline in CH construction, while maintaining competitive query performance and CH graph size.
We released our code and all datasets used in this paper.
\end{abstract}





\renewcommand\footnotetextcopyrightpermission[1]{} 
\fancyhead{} 

\maketitle

\section{Introduction}

Computing shortest distance is one of the most fundamental graph problems,
playing a vital role in various applications, such as navigation on road networks.
To accelerate point-to-point distance queries, many existing solutions use two-phase approaches, which
preprocess the graph and construct an index---often an auxiliary graph---to facilitate queries.
Among the most notable two-phase solutions is the \emph{Contraction Hierarchies (CH)}~\cite{geisberger2008contraction}, which is mainly designed for sparse networks such as road networks, and is widely used in practice, for instance in Google Maps.
In addition to being used on its own, CH is also a vital component in other approaches for various applications on distance queries, such as Transit Node Routing~\cite{arz2013transit,bast2006ultrafast,bast2007fast}, Hub-Based Labeling~\cite{abraham2011hub,abraham2012hierarchical,delling2013hub}, and some renowned algorithms including \CHASE~\cite{bauer2010combining} and \PHAST~\cite{delling2013phast}.
We refer the audience to the excellent surveys~\cite{sommer2014shortest, bast2016route, madkour2017survey} for more background of the state-of-the-art techniques for route planning.

\begin{table}[t]
  \small
  \centering
    \begin{tabular}{c|c|c}
    \multicolumn{1}{c|}{Algorithms} & \multicolumn{1}{c|}{Preprocessing time} & \multicolumn{1}{c}{Query time} \\
    \midrule
    \multicolumn{1}{c|}{Dijkstra~\cite{dijkstra1959}} & -     & 7.20 s \\
    \multicolumn{1}{c|}{$\Delta$-stepping$^*$~\cite{dong2024pasgal}} & -     & 0.320 s \\
    \midrule
    \multicolumn{1}{c|}{\RoutingKit~\cite{RoutingKit,dibbelt2016customizable}}    & 2466 s & 79.1 $\mu$s \\
    \multicolumn{1}{c|}{\PHAST~\cite{PHAST,delling2013phast}} & 1341 s & 138 $\mu$s \\
    \multicolumn{1}{c|}{\chconstructor$^*$~\cite{chconstructor2024code}}    & 1527 s & 317 $\mu$s \\
    \multicolumn{1}{c|}{\OSRM$^*$~\cite{OSRM,luxen2011real}}  & $307$s & 163 $\mu$s \\
    \multicolumn{1}{c|}{ \textbf{Ours$^*$}}  & \textbf{23.1 s} &  \textbf{93.3 $\mu$s} \\
    \end{tabular}%
  \caption{\textbf{Preprocessing and point-to-point shortest path query time on the road network North America~\cite{roadgraph}.}
  The graph has 87 million vertices and 113 million edges.
  1 $\mu$s = $10^{-6}$ s.
  (*) denotes parallel implementations; others are sequential.
  The reported query time is the average across $1000$ randomly selected pairs.
  Dijkstra~\cite{dijkstra1959} and $\Delta$-stepping~\cite{dong2024pasgal} are sequential and parallel SSSP algorithms for comparison.
  }
  \label{tab:teaser}%
\end{table}%

\begin{figure}[!t]
  \centering
  \small
  \includegraphics[width=\columnwidth]{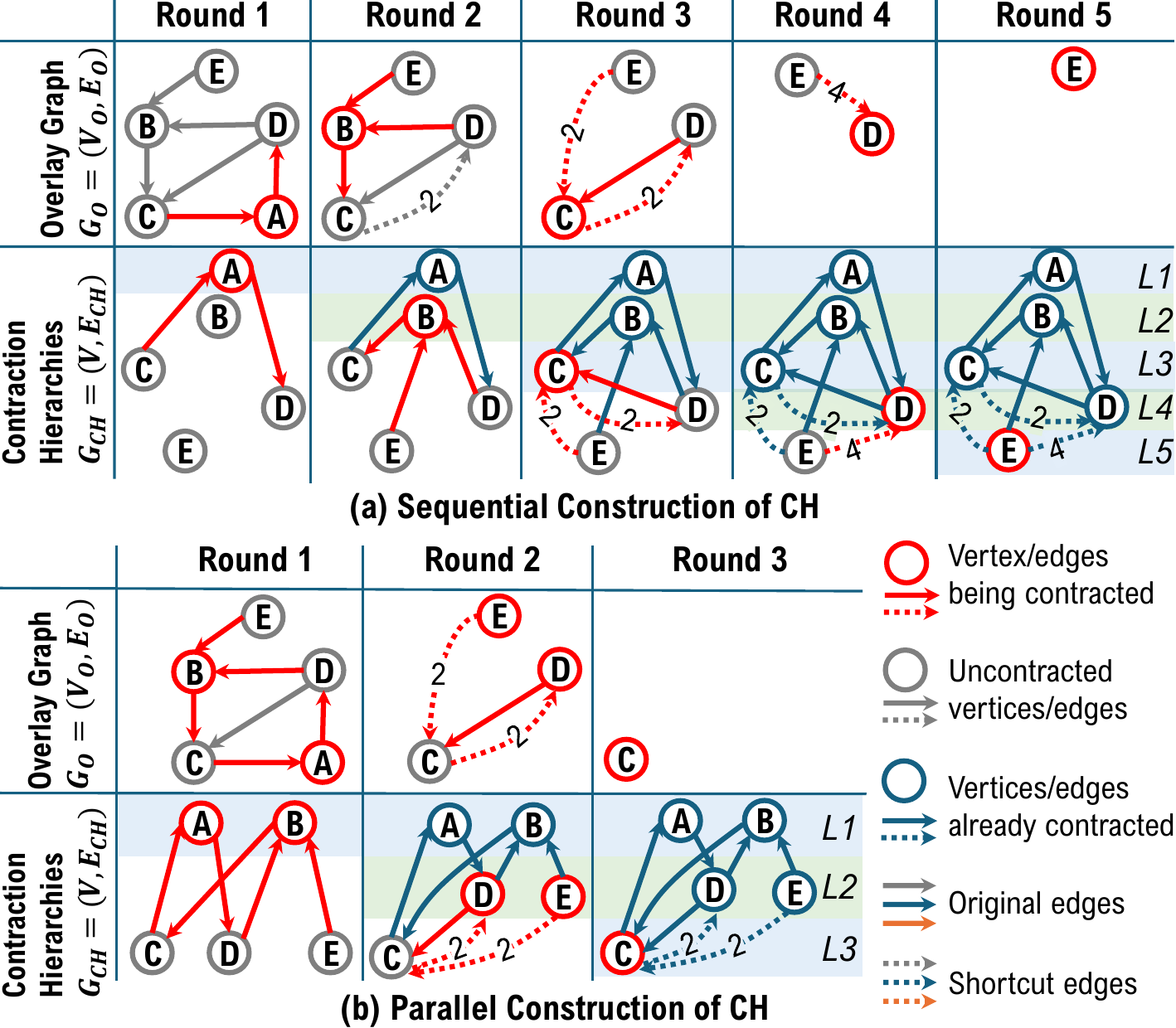}
  \caption{\textbf{Illustration of the construction of Contraction Hierarchies.}
  For simplicity, we assume the input graph has unit weights, omitting the weight
  ``1'' from the graph.
  }\label{fig:ch}\label{fig:pch}\label{fig:ch-example}
\end{figure}

In this paper, we consider a graph $G=(V,E)$ with an edge weight function $w:E\mapsto \mathbb{R}^{+}$.
Given two vertices $s,t\in V$, a distance query asks for the shortest distance from $s$ to $t$.
The idea of CH, as illustrated in \cref{fig:ch-example}(a), is to ``contract'' the graph into a hierarchy, which is an auxiliary graph $\PCH$ that preserves the shortest distances in $G$, such that
distance queries on $\PCH$ can be much faster.
In each iteration, one vertex $u$ and all its edges are moved from the original graph to an auxiliary graph $\PCH$, forming a \emph{level} in $\PCH$.
Additionally, extra edges (\defn{shortcuts}) are added to the original graph, forming the \defn{overlay graph} $\overlayG$, to preserve the pairwise distances among the remaining vertices.
Finally, all vertices in $\PCH$ form the \defn{contraction hierarchies (CH)}.
A distance query requires a bidirectional search on $\PCH$.

Empirically, a distance query on the CH touches far fewer vertices than in the original graph, thereby resulting in much faster query speed.
As shown in \cref{tab:teaser}, for the North American roadmap from OpenStreetMap~\cite{roadgraph}, queries can be $10^3$--$10^5$ faster than directly using SSSP algorithms, such as Dijkstra~\cite{dijkstra1959} and $\Delta$-stepping (a parallel SSSP algorithm)~\cite{meyer2003delta,dong2021efficient}.
However, this impressive querying speed comes at a cost---\emph{constructing $\PCH$ is expensive}.
Existing sequential solutions use 22--41 minutes, and the best parallel solutions need more than 300 seconds.
Such long preprocessing time may require significant computational resources, and limit its adaptability to large graphs.
Such running time is reasonable for a sequential algorithm---both \RoutingKit~\cite{dibbelt2016customizable} and \PHAST~\cite{delling2013phast} take $100\times$ construction time than Dijkstra, which is acceptable considering the speedup for queries.
However, the existing parallel CH algorithms are only $4.4\times$ faster than (sequential) \PHAST{} even on a 96-core machine.
The results indicate a significant gap,
suggesting the potential for improvements in parallelism in CH construction algorithms.

In this paper, we propose \defn{\oursys{}} (\textbf{S}calable \textbf{P}arallelization \textbf{o}f \textbf{C}ontraction \textbf{H}ierarchies),
which supports \textbf{scalable and efficient contraction hierarchies construction with high parallelism} without compromising the query performance.
The performance gain is mainly from good parallelism and algorithmic improvements for sequential performance.

\hide{
The trade-off for such a fast query speed is a slow preprocessing time.
Generating a CH is computational expensive, mostly due to deciding which edges should be added back to the overlay graph $\overlayG$ in each round (details in \cref{sec:original_ch}).
Empirically, \RoutingKit~(RK)~\cite{RoutingKit,dibbelt2016customizable}, a SOTA sequential CH solution, uses times around hundreds of running times for Dijkstra on the same graph, which is tolerable considering the speedup for the queries.
However, unlike Dijkstra that has highly parallel counterparts, existing parallel CH algorithms suffer from severe scalability issues, making them almost no faster than the sequential ones even on a 96-core machine.
}

Conceptually, parallelizing CH construction is not hard.
In 2009, Vetter~\cite{vetter2009parallel} pointed out that vertices in an independent set (i.e., vertices that do not share edges) can be contracted in parallel, as illustrated in \cref{fig:ch-example}(b).
However, despite being studied in many later papers~\cite{OSRM,luxen2011real,karimi2019gpu,karimi2020fast,kieritz2010distributed,chconstructor2024code},
we are unaware of scalable multicore CH implementations.
In fact, constructing CH is a sophisticated process with intensive computation.
Achieving good performance requires high parallelism in all steps and careful algorithmic design.
We observe two major challenges.
First, identifying the vertices to contract involves simulating contractions on many (if not most) vertices,
which calculates distances for numerous vertex pairs.
Such a process is expensive, requiring high parallelism to enable good performance.
The second challenge is to dynamically maintain the graph in parallel,
since the graph may experience rapid changes, such as the removal of contracted vertices/edges and the insertion of shortcuts.
Existing solutions may sacrifice parallelism (e.g., using locks) to support such dynamic updates.
Ideally, a scalable solution would necessitate graph maintenance support with full parallelism.

\myparagraph{Contributions.}
We propose \oursys{}, which consists of a new algorithmic framework for CH, and a high-performance implementation.
Our improvement lies in both \emph{algorithm} and \emph{data structure} design with high parallelism.
Algorithmically, we introduce the new \step{LocalSearch} step in construction to address the high cost of simulated contraction on a large fraction of vertices.
This step gathers all the pairs of vertices for distance computation, and processes them in a batch in parallel.
Processing them as a batch allows for combining and eliminating repeated computations, offering high potential to exploit parallelism, as well as
a \emph{pruning} process to remove suboptimal shortcuts added in previous rounds.
Finally, \oursys{} memoizes the computed distances in this step to facilitate later computation.
We provide more information in \cref{sec:prune}.
Combining these benefits, the \step{LocalSearch} step improves both the sequential running time and parallelism.

Another key performance improvement in \oursys{} comes from using efficient parallel \emph{data structures}.
As mentioned, to get a highly parallel solution, we need to efficiently handle edge updates to the graph in parallel.
We use the \emph{phase-concurrent hash table}~\cite{shun2014phase} to buffer newly added shortcuts in each round, which allows for efficient lock-free concurrent insertions of new edges.
However, combining these edges to the overlay graph in each round requires scanning the entire hash table array and the edge list of the overlay graph. This may be inefficient in most rounds where only a small number of shortcuts are added.
Our solution employs a lazy update scheme to delay the combination of the shortcuts to the overlay graph,
while still making the delayed shortcuts visible to future computations.
We introduce this technique in \cref{sec:ds}.
This approach reduces the cost of maintaining the graph from $39.9\%$ to $19.8\%$ of the overall running time on average across all tested graphs.

\hide{
\begin{figure*}[t]
  \centering
  \begin{subfigure}[t]{.45\textwidth}
  \includegraphics[height=0.4\textwidth]{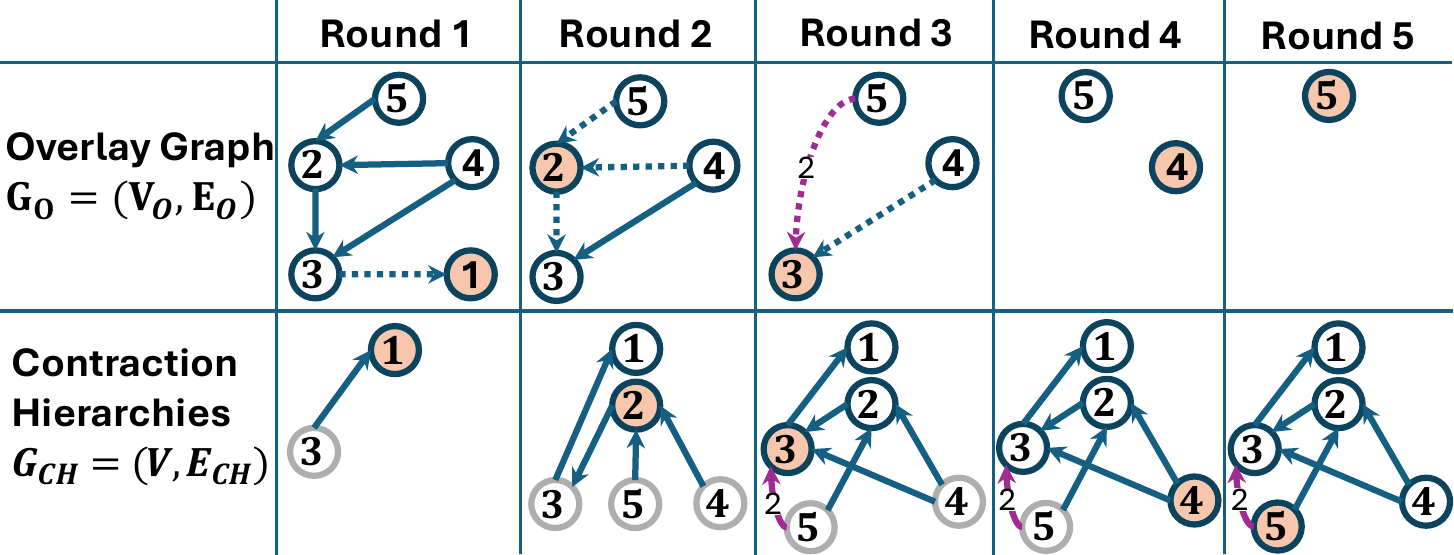}
  \caption{\textbf{Illustration for the Construction of Contraction Hierarchies.}
  For simplicity, we omit the edge weights from the graph
  and assume that vertices are contracted in the order of their vertex IDs.
  }
  \end{subfigure}%
  \hspace{0.05\textwidth} 
  \begin{subfigure}[t]{.45\textwidth}
  \includegraphics[height=0.4\textwidth]{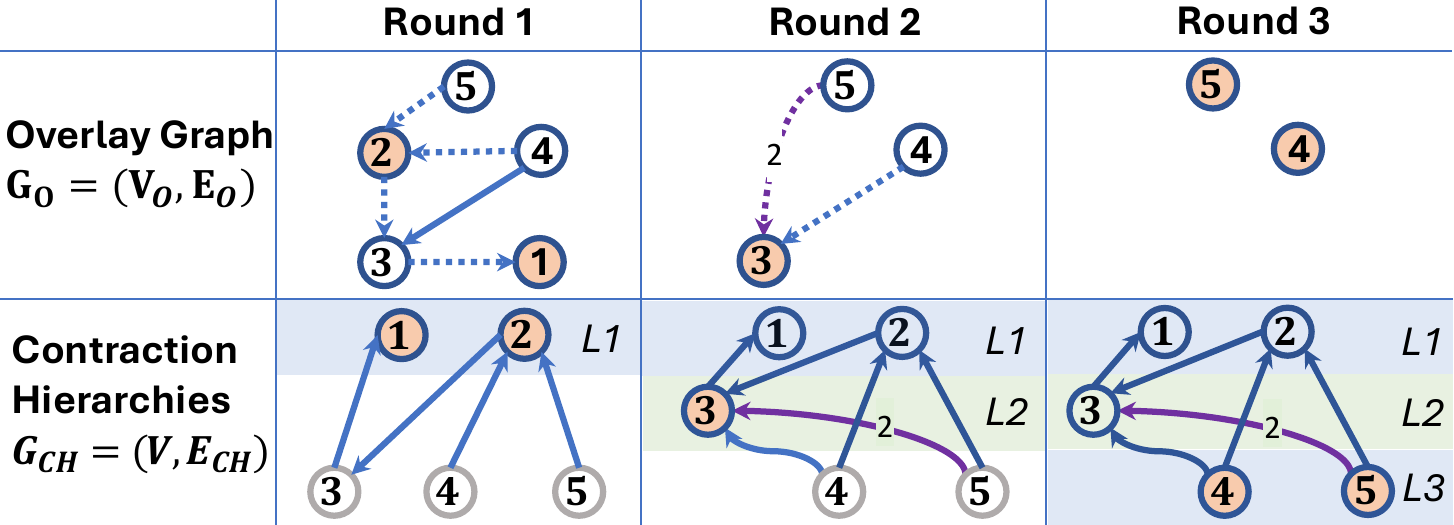}
  \caption{\textbf{Illustration for the Parallel Construction of Contraction Hierarchies.}
  See~\cref{fig:ch} for the sequential construction.
  In each round, a vertex $v$ can be contracted in parallel if $\priority[v] < \priority[u]$ for $u\in \n(v)$.
  }\label{fig:pch}
  \end{subfigure}%
  \caption{
  The vertex being contracted is highlighted in orange.
  Edges incident to the contracted vertex that are moved from $\overlayE$ to $\ECH$ are shown as dashed lines.
  The shortcuts introduced by the algorithm are highlighted in purple.
  }
\end{figure*}
} 

We compare \oursys{} with the state-of-the-art sequential and parallel solutions on various graphs.
Even the sequential running time of \oursys{} is as fast as the sequential baselines.
Due to our new design, \oursys{} achieves high parallelism and good scalability with large numbers of processors (see \cref{fig:scalability}).
On North America in \cref{tab:teaser}, \oursys{} is $13.3$--$107\times$ faster in construction than the baselines while achieving similar query performance.
On the 16 graphs we tested in \cref{tab:ch_algo_cmp}, compared to the \emph{fastest baseline on each graph},
\oursys{} is 3.83--41.0$\times$ faster, with an average of $8.85\times$.
We also conduct in-depth performance studies of the CH algorithms in \cref{sec:exp}.
We released our code in~\cite{chcode}.
\myparagraph{Preliminaries.}
We focus on the shared-memory setting with fork-join parallelism. 
The computation starts with one thread. 
A thread can \forkins{} two child software \thread{s} to work in parallel.
When both children complete, the parent thread continues.
Such a computation can be executed by a randomized work-stealing scheduler~\cite{blumofe1999scheduling,arora2001thread,blelloch2020optimal}. 
We use the atomic operation \cas{}$(p,v_{{\scriptsize\mbox{\it old}}}, v_{{\scriptsize\mbox{\it new}}})$, which checks if the memory location pointed to by $p$ has the value $v_{\scriptsize\mbox{\it old}}$, 
and if so, changes the value to $v_{\scriptsize\mbox{\it new}}$. The function returns \true{} if it successfully changes the value, and \false{} otherwise. 
We also use \WriteMin{}$(p,\mathit{v})$, which reads the memory location pointed to by $p$ and writes value $v$ to it if $v$ is smaller than the current value.

We consider a weighted graph $G=(V,E,w)$ with $n=|V|$, $m=|E|$, and an edge weight function $w:E\to \R^+$.
For $v\in V$, we define $\nin(v)=\{u\,|\,(u,v)\in E\}$ as the \defn{in-neighbors} of $v$,
and $\nout(v)=\{u\,|\,(v,u)\in E\}$ as the \defn{out-neighbors} of $v$.
We use $\n(v)=\nin(v)\cup \nout(v)$. 
For a set of vertices $V'\subset V$, we define $\n(V')$ as all neighbors of vertices in $V'$.
Note that the main process of the CH construction algorithm is to update the \emph{overlay graph} $G_O$ by
removing contracted vertices and incident edges, and inserting shortcuts. 
Therefore, when we use $\nin(\cdot),\nout(\cdot),\n(\cdot)$ in the algorithm description, 
we \textbf{always refer to the neighborhood of a vertex on the \emph{overlay graph}. }

\begin{table}[t]
  \small
  \centering
    {\fontsize{8}{9.5}\selectfont\begin{tabular}{@{}ll@{}}
    \toprule
    $G=(V,E)$&: The input graph \\
    ${\overlayG}=(\overlayV,\overlayE)$&: The overlay graph \\
    ${w(u,v)}$ &: Weight for edge $(u,v)$\\
    ${\priority[v]}$ & : The score of vertex $v$\\
    ${\nin(v)}$& : In-neighbors of vertex $v$ \\
    ${\nout(v)}$& {: Out-neighbors of vertex $v$} \\
    ${\n(v)}$& {: In- and out-neighbors of vertex $v$} \\
    \multicolumn{2}{l@{ }}{(Unless otherwise specified, we use $\nin(v)$, $\nout(v)$ and $\n(v)$ to refer}\\
    \multicolumn{2}{l@{ }}{ to the neighbors of $v$ on the \textbf{overlay graph})}\\
    \bottomrule
    \end{tabular}}%
    \caption{General notations used for contraction hierarchies. }
  \label{tab:notation-table}%
\end{table}%





\section{Contraction Hierarchies}\label{sec:original_ch}

Contraction Hierarchies (CH) was proposed by Geisberger et al.~\cite{geisberger2008contraction} in 2008, building upon the concept of highway hierarchies~\cite{sanders2006engineering, knopp2007computing}.
CH is one of the most widely adopted techniques in route planning and is used in many systems, such as Google Maps.

Let $G=(V,E)$ be the input graph. Its \defn{Contraction Hierarchies} {$\boldsymbol{\CH=(V,\ECH)}$} is an auxiliary graph that preserves the pairwise distances of $G$.
Distance queries on $\CH$ can be orders of magnitude faster than on $G$ itself, particularly for sparse graphs such as road networks.
The CH algorithm iteratively contracts the ``least important'' vertex (determined by a user-specified score function) from the graph, and adds its edges to the contraction hierarchy. Additional edges (shortcuts) are added to the remaining graph (referred to as the overlay graph $\overlayG$) to preserve the distances.
Each contracted vertex represents a \emph{level} in the CH, forming a hierarchy.
This process is illustrated in \cref{fig:ch-example}(a).
A distance query from $s$ to $t$ is performed by running a bidirectional search from both $s$ and $t$ on the CH.
The query algorithm is presented in \cref{app:query}.
By contracting vertices and adding shortcuts, the hop distances between vertices are greatly reduced, allowing for much faster query performance.

\hide{
The high-level idea behind CH is that, most real-world networks contain significant nodes that appear on many shortest paths, and insignificant ones vice versa.
Taking roadmaps as examples, significant nodes can be highways and their entrances, while insignificant ones are local connections.
In a CH, insignificant vertices are contracted and removed in earlier rounds, minimally affecting the graph and rarely appearing in most searches.
After they are removed, the graphs are simplified, and shortcuts are added to connect significant vertices.
This will largely accelerate the query performance by 1) visiting fewer vertices due to the simplification of the graphs, and 2) finding the shortest paths in fewer hops by using the shortcuts.
Unlike the previous approaches, CH builds a full hierarchy for $n$ levels, which only needs to find the \emph{least} useful vertex at a time, instead of clustering vertices into various levels of significance.
This approach greatly simplifies the algorithm design.
}

Given its wide range of applications~\cite{funke2019pathfinder,funke2015placement,abeywickrama2016k,sanchez2022simulation,ouyang2020efficient}, CH has been extensively studied.
In this section, we first review the algorithm proposed by Geisberger et al.~\cite{geisberger2008contraction} along with the fundamental concepts.
Then in \cref{sec:exist-para-ch}, we introduce existing parallel solutions.
Other related studies are discussed in \cref{sec:related}.

\begin{algorithm}[t]
  \small
  \caption{Contraction Hierarchies Construction~\cite{geisberger2008contraction}\label{alg:ch}}
  \SetKwInOut{Maintains}{Maintains}
  \KwIn{A graph $G=(V,E)$.}
  \KwOut{The contraction hierarchy $\CH=(V,\ECH)$.}
  \Maintains{The overlay graph $\overlayG=(\overlayV,\overlayE)$\\
  The score of each vertex $\priority[\cdot]$}
  \DontPrintSemicolon

  $\overlayG\gets G, \ECH\gets \emptyset$\\ \label{line:och-init}
  Compute the score $\priority[\cdot]$ for all $v\in V$\\ \label{line:och-init-score}
  \While{$\overlayV \neq \emptyset$}{ \label{line:och-finish}
    \tcp{Select the vertex $v$ with lowest score from $\overlayV$}
    $v=\argmin_{v'\in \overlayV}\priority[v']$\label{line:och-select}
    \For{$u_1\in \nin(v)$\label{line:simulated-start}}{
      \For{$u_2\in \nout(v)$} {
        \tcp{Shortest path distance from $u_1$ to $u_2$}
        $l \gets \fname{WPS}(u_1,u_2)$\label{line:och-wps} \\
        \If{$w(u_1,v)+w(v,u_2)=l$}{ \label{line:och-check}
          $\overlayE = \overlayE \cup \{(u_1, u_2)\}$ \\ \label{line:och-addedge}
          $w(u_1,u_2) = w(u_1,v)+w(v,u_2)$ \label{line:och-weight}
        }
      }
    }
    $\AffectedNeighbors \gets \n(v)$ \\ 
    Remove $v$ and all its edges from $\overlayG=(\overlayV,\overlayE)$ and add them to $\CH$  \label{line:och-moveedge}\\
    Update the scores $\priority[u]$ for all $u\in\AffectedNeighbors$ \label{line:och-rescore}


  }
  \Return {$\CH$}
\end{algorithm} 

\subsection{Sequential Solutions}\label{sec:och-construct}

We present the sequential construction algorithm for CH in Alg.~\ref{alg:ch}.
The algorithm takes a graph $G=(V,E)$ as the input, and computes the CH for $G$ in  $\CH=(V,\ECH)$.
As mentioned, the contraction process requires removing (contracting) vertices, and adding shortcuts to preserve the shortest distances.
To avoid destroying the input graph, an \emph{overlay graph} $\overlayG=(\overlayV,\overlayE)$ is used to reflect the changes.
A score array $\priority[\cdot]$ is also maintained for all uncontracted vertices,
representing their ``importance'' and deciding the contracting order.

The algorithm starts by computing the initial scores of all vertices (line~\ref{line:och-init-score}).
Then, it iteratively selects and contracts the vertex with the lowest score (line~\ref{line:och-select}) until the overlay graph is empty (line~\ref{line:och-finish}).
When contracting a vertex $v$,
the algorithm computes the shortest paths (line~\ref{line:och-wps}) from each $u_1\in \nin(v)$ to each $u_2\in\nout(v)$ by running Dijkstra's algorithm. These shortest path queries are referred to as the \emph{Witness Path Search (WPS)}~\cite{geisberger2008contraction,karimi2019gpu}.
If the shortest path length $l$ from $u_1$ to $u_2$ is the same as $w(u_1,v)+w(v,u_2)$ (line~\ref{line:och-check}), then $v$ can be on the shortest path from $u_1$ to $u_2$.
Thus, the shortcut edge $(u_1,u_2)$ with weight $l$ is added to $\overlayE$ to preserve the distances on $\overlayE$ (line~\ref{line:och-addedge} and \ref{line:och-weight}).
After that, $v$ and all its incident edges are removed from the overlay graph $\overlayG=(\overlayV,\overlayE)$,
and are added to the CH $\CH$.
Contracting $v$ may cause the score of all its neighbors to change, which are updated on line~\ref{line:och-rescore}.
Finally, the algorithm moves on to contract the next vertex, until all vertices are contracted.


\myparagraph{Vertex Score and Edge Difference.}
Scoring the vertices is the most time-consuming part of CH construction and significantly impacts the shortest path query performance in CH.
The most widely used score function is the \emph{edge difference}~\cite{geisberger2008contraction}.
For a vertex $v$, the edge difference is the change in the number of edges on the overlay graph $\overlayG$ after contracting $v$,
i.e., the number of shortcuts added minus the number of edges incident to $v$.
A vertex with a higher edge difference is considered more important because its contraction has a larger impact on the graph's structure.
To compute the number of shortcuts to be added, a \defn{simulated contraction} on $v$ is required,
which virtually generates and counts the shortcuts.
This simulation is very similar to line~\ref{line:simulated-start}--\ref{line:och-weight} except it does not add edges or modify edge weights in $\overlayG$.

In existing work~\cite{geisberger2008contraction,geisberger2012exact}, the score function is usually defined as a weighted combination of edge difference and many other metrics, such as vertex degree, hop distance, and hierarchy depth.
Among them, edge difference has the highest weight and is usually the most computationally expensive part.
Therefore, for simplicity, our algorithm description uses edge difference as the score function.
Our algorithmic ideas are independent of the score function.

\subsection{Existing Parallel Algorithms}\label{sec:exist-para-ch}

The construction for CH is very costly, so it is natural to consider parallelizing this process by contracting many vertices in parallel.
However, note that we cannot contract arbitrary vertices in parallel.
For instance, if two adjacent vertices $u$ and $v$ are contracted together, $u$ may choose to add a shortcut incident to $v$.
Since $v$ is removed from the overlay graph at the same time, the shortcut may not be added correctly, resulting in distances not being preserved accurately on the overlay graph.
In 2009, Vetter~\cite{vetter2009parallel} first observed that vertices that do not share edges (i.e., an \emph{independent set}) can be contracted in parallel, and proposed the first parallel CH construction algorithm.
In this case, all added shortcuts will be connected to uncontracted vertices and are thus safe to add to the overlay graph.

Almost all later parallel solutions (e.g.,~\cite{karimi2020fast,chconstructor2024code,luxen2011real,kieritz2010distributed,karimi2019gpu}) follow Vetter's high-level idea.
In each round, an independent set (IS) of vertices is identified to be contracted.
All these vertices will be placed on the same level in the CH.
After that, shortcuts will be added normally.
This process is repeated until all vertices have been contracted.
As with the sequential algorithm that first contracts vertices with the lowest score, the vertices in the IS should also have low scores in general.
In Vetter's original algorithm, the IS includes all vertices with the smallest score within its $k$-hop neighborhood.
\osrm{}~\cite{OSRM,luxen2011real} specifically uses $k=2$. Our implementation also follows this idea and uses $k=1$.
An existing GPU algorithm~\cite{karimi2019gpu} and \chconstructor{}~\cite{chconstructor2024code} find a maximal independent set (MIS) and include all vertices in this MIS with scores smaller than a threshold.

\hide{
For instance, if $u$ and $v$ share an edge and are contracted together,
$u$ will add shortcuts to $v$ and $u$'s other neighbors to preserve the distances in the overlay graph (and vice versa), but since $v$ will also be removed from the graphs, the shortcuts will not finally included, causing the correctness issue.

However, one can verify that it is safe to contract a batch of nodes that do not share any edge, which is algorithmically referred to as an independent set.
Indeed, Vetter observed this in 2009~\cite{vetter2009parallel}, the year after CH was proposed.
Vetter's solution is quite straightforward.
In each round, it picks a batch of vertices to contract, each has lower score than its neighbors' scores.
This will guarantee that the selected vertices form an independent set.
Then, vertices are contracted in parallel, so as updating overlay graph and the vertex scores after contraction.
The algorithm runs in rounds until all vertices are contracted.
}

\myparagraph{Challenges to Achieve High-Performance Implementations.}
While Vetter's work reveals the potential parallelism in the algorithm, the idea itself is not sufficient to enable a high-performance solution.
Many challenges remain in other parts of the algorithm.
We highlight two components here.
The first part is the process to (re)compute the scores of vertices.
When multiple vertices are contracted together, a large number of simulated contractions is needed, which all involve running WPSes by Dijkstra's algorithm.
The second challenge is to update the overlay graph in parallel, since
contracting vertices in parallel results in a bulk of new shortcut edges that need to be added to the overlay graph.
Both parts (performing WPS and updating the graph) involve expensive computation, and require careful
algorithm design and parallelization.
In our experiments, we test two existing parallel solutions \chconstructor~\cite{chconstructor2024code} and \OSRM~\cite{OSRM,luxen2011real} based on Vetter's algorithm.
In tests on a 96-core machine with 16 graphs (\cref{tab:ch_algo_cmp}), their performance is only up to $7.04\times$ faster than a highly-optimized sequential implementation, and can be up to 3.59$\times$ slower on some graphs.

In the next section, we present our solution that achieves a scalable and efficient parallel CH construction algorithm.

\hide{
Many challenges that cause scalability issues still remain unsolved.
We highlight two of them here.

The first challenge is the \emph{load imbalance} among contracting vertices in the same batch.
The most costly part during the contraction of a vertex is the simulated contraction, and the running times for different vertices in the batch varies greatly.
This may sound counterintuitive since vertices on sparse networks (e.g., road networks) have similar degrees, leading to about the same number of WPS searches.
However, some (important) vertices can quickly gather more neighbors though the shortcuts during the contraction, causing more WPS searches.
These searches are usually more costly since they touch the denser regions in the graph.
Due to all these reasons, the cost to contract some vertices may take time 3--4 \yan{check} orders of magnitude more than others.
Directly parallelizing the contraction of each vertex will cause significant idle times for other processors waiting for the slowest one.

Another challenge is to support \emph{concurrent edge updates} (insertions and deletions) to both the CH and the overlay graph (see \cref{fig:pch}).
Existing parallel solutions just leave this part completely sequential, whereas a scalable solution would necessitate parallel support for these operations.
}

\hide{

Although CH is helpful in improving the query time by orders of magnitudes, its construction phase is time-consuming.
Therefore, seeking parallel solution for CH construction is very important.
However, CH construction is hard to be parallelized, since if two vertices $v_1,v_2$ to be contracted are adjacent,
conflicts can happen due to contention.
As introduced in~\cref{sec:original_ch}, during the construction of CH, the contraction of a vertex may involve inserting shortcut edges between its neighbors to preserve the correctness of the shortest path distances in $\overlayG$.
Therefore, two adjacent vertices cannot be contracted in parallel, as shortcut edges cannot be inserted to contracted vertices.

\begin{algorithm}
  \small
  \caption{Vetter's Parallel Contraction Hierarchies Construction Framework~\cite{vetter2009parallel}} \label{alg:vch}
  \SetKwFor{parForEach}{ParallelForEach}{do}{endfor}
  \KwIn{A graph $G=(V,E)$.}
  \KwOut{The contraction hierarchy $\PCH=(V,\ECH)$}
  \SetKwInOut{Maintains}{Maintains}
  \Maintains{The overlay graph $\overlayG=(\overlayV,\overlayE)$\\
  \DontPrintSemicolon
  The score of each vertex $\priority[\cdot]$}
      $\overlayG\gets G, \ECH\gets \emptyset$\label{line:vch-init} \\
      $\AffectedNeighbors \gets V$\label{line:vch-initscore}

      \While{$\overlayV \neq \emptyset$}{\label{line:vch-stop_cond}
          \mbox{\upshape \textsf{Score}$(\AffectedNeighbors,\overlayE)$}\tcp*{update the scores of vertices in $\AffectedNeighbors$}\label{line:vch-scoring}
          $\FeasibleCandidates\gets$\mbox{\upshape \textsf{Select}$(\overlayV,\overlayE)$}\tcp*{select an independent node set $\FeasibleCandidates$ with low scores}\label{line:vch-selecting}
          $\AffectedNeighbors\gets\n(\FeasibleCandidates)$\tcp*{find $\AffectedNeighbors$, the set of neighboring vertices of $\FeasibleCandidates$ that require score update in the next iteration}\label{line:vch-findneighbor}
          \mbox{\upshape \textsf{Contract}$(\FeasibleCandidates,\overlayE,\ECH)$}\tcp*{contract $\FeasibleCandidates$ in parallel}\label{line:vch-contracting}
          $\overlayV\gets \overlayV-\FeasibleCandidates$\label{line:vch-remove}
      }

      \Return{$\CH=(V,\ECH)$}


\end{algorithm}

\myparagraph{Vetter's Parallel Contractoin Hierarchies Framework.}
Vetter proposed the first parallel contraction hierarchies algorithm.
Its algorithm framework is presented in~\cref{alg:vch}, which is widely adopted by many existing studies~\cite{funke2017personal,luxen2011real} on parallel construction of CH.
To overcome the challenge that adjacent vertices cannot be contracted in parallel,
Vetter~\cite{vetter2009parallel} proposed a method based on \defn{independent sets (IS)}.
Specifically, in each iteration of the contraction, one can simultaneously contract an independent set (IS) of vertices $\FeasibleCandidates$ from the overlay vertices $\overlayV$, such that every pair of vertices $v_1,v_2\in\FeasibleCandidates$ are not adjacent.
This is feasible because the contraction of $\FeasibleCandidates$ only introduce new shortcut edges inserted between the remaining vertices.

However, contracting multiple vertices in one iteration brings in a new problem ---
as discussed in~\cref{sec:och-construct}, the good query performance highly relies on node ordering given by the scores.
To make sure that vertices in $\FeasibleCandidates$ have low scores compared to the rest of the vertices,
Vetter's algorithm only chooses vertices with the lowest importance within its $k$-neighborhood
(the set of vertices reachable with at most $k$ hops).

Similar to the sequential CH algorithm, Vetter's algorithm also maintains the overlay graph $\overlayG=(\overlayV,\overlayE)$,
and starts by initializing $\overlayG$ to $G$ and $\ECH$ to an empty set (line~\ref{line:vch-init}).
In each iteration, it maintain a set of vertices $\AffectedNeighbors\subseteq \overlayV$ that needs a score update.
At the beginning, $\AffectedNeighbors=V$ since the scores of all vertices need to be initialized (line~\ref{line:vch-initscore}).
After initialization, Vetter's algorithm runs in iterations.
In each iteration, it first update the scores of the vertices in $\AffectedNeighbors$ (line~\ref{line:vch-scoring}).
Based on the latest scores, it selects an independent set $\FeasibleCandidates$ with low scores from $\overlayV$ (line~\ref{line:vch-selecting}) set and its neighboring vertices as $\AffectedNeighbors$ (line~\ref{line:vch-findneighbor}).
$\AffectedNeighbors$ is set to the neighbors of $\FeasibleCandidates$ since their incident edges could have changed due to new shortcuts,
and thus their scores need to be recomputed.
The vertices in $\FeasibleCandidates$ will be contracted (line~\ref{line:vch-contracting}),
which involves moving their incident edges from the overlay graph to the CH and adding pairwise shortcuts from in- to out-neighbors, as in the sequential CH construction.
Finally, $\FeasibleCandidates$ is removed from $\overlayV$ (line~\ref{line:vch-remove}), followed by the next iteration of contraction, or until all vertices have been contracted (line~\ref{line:vch-stop_cond}).

The first issue is the redundant execution of WPSes and poor load balancing of the simulated contraction process.
As introduced in~\cref{sec:original_ch}, simulated contraction is used for both score calculation during node ordering and vertex contraction.
In each round of parallel construction of $\PCH$, we contract a large number of vertices $\FeasibleCandidates$ in parallel instead of one in the sequential setting, which significantly increases the number of vertices requiring score updates $\AffectedNeighbors$.
\xiaojun{Parallelizing CH construction does not increase $\Sigma |\AffectedNeighbors|$ (total number of updates). In fact, it reduces it comparing to the sequential algorithm.}
\zijin{Here we are discussing $\AffectedNeighbors$ in each round, not in total}
In previous parallel implementations, each simulated contraction procedure is assigned to a separate thread, with the WPSes within each procedure being executed sequentially.
This approach is very time-consuming and lacks parallelism, suffering from poor load balancing, as many WPSes are redundantly executed multiple times in parallel, and the cost of each simulated contraction varies significantly.
Our algorithm avoids redundant WPS execution by storing the results of WPSes in a parallel hashmap and addresses the load imbalance issue by decoupling WPSes from simulated contraction.
More details are provided in~\cref{sec:prune}.

The second issue is the lack of parallelization in the insertion of shortcut edges.
Contracting $\FeasibleCandidates$ in parallel means that shortcuts $\ShortcutEdges$ are being added concurrently.
However, in previous parallel implementations [], while vertex contraction is parallelized, shortcut insertion remains sequential.
Shortcuts are inserted into a buffer array, with a mutex used to avoid conflicts in critical sections.
After finishing all the contractions, the array is sorted and merged back into the compressed sparse row (CSR) for storing overlay edges $\overlayE$.
\xiaojun{Either define CSR in prelim or remove it.}
This method is functional but lacks parallelism.
Our algorithm uses a modified parallel hashmap to store these shortcuts, which supports both parallel insertions and queries.
More details are provided in~\cref{sec:node_contractions}.
} 
\hide{
\begin{figure}[t]
  \small
  \centering
  \includegraphics[width=\columnwidth]{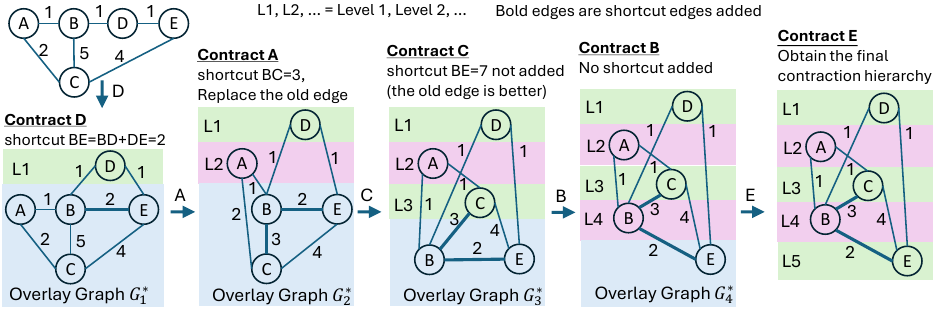}
  \caption{Sequential Contraction Hierarchy}\label{fig:ch-seq}
\end{figure}
}

\hide{

by iteratively contracting vertices and adding shortcuts.
On the CH, one can answer the point-to-point shortest path queries efficiently,
with orders of magnitudes speedup over a standard Dijkstra on road networks.
Some single-source shortest paths (SSSP) solutions on road networks, such as \Phast~\cite{delling2013phast}, is also based on CH.

In real-world point-to-point route navigation, the importance of roads in the shortest path typically increases as one moves farther from the source and reaches major highways, then decreases as the route approaches the destination and transitions back to local roads.
The CH algorithm, consisting of the construction phase and the query phase, utilizes this property.

In the construction phase, vertices are categorized by their level of importance and shortcuts are added to preserve the correctness of the shortest paths.
In the query phase, a bidirectional search is performed, with one search goes forward from the source and one search goes backward from the target.
Both searches only lead to more important vertices in the hierarchy and eventually meet at the most important vertex along the shortest path between them.
By bypassing relatively unimportant and irrelevant vertices, the cost of CH query is significantly reduced.
The survey by Sommer~\cite{sommer2014shortest} shows that its query performance dominates in the regime of low space overhead route planning algorithms.
}

\hide{
The Contraction Hierarchies (CH) algorithm \cite{geisberger2008contraction} is a preprocessing technique that enables much faster point-to-point shortest-path query on road networks.
The construction of \defn{Contraction Hierarchies $\boldsymbol{\CH=(V,\ECH)}$} consists of two main parts: ordering the vertices in ascending order of importance, and iteratively contracting the vertices in this order.
During the contraction, an \defn{overlay graph $\boldsymbol{\overlayG=(\overlayV,\overlayE)}$}, which contains all remaining vertices and edges, is dynamically maintained.
}

\section{Our Parallel CH Construction Algorithm}\label{sec:contraction}

\begin{algorithm}
    \small
    \caption{Our Parallel Algorithm for CH Construction} \label{alg:pch}
    \DontPrintSemicolon
    \SetKwFor{parForEach}{ParallelForEach}{do}{endfor}
    \KwIn{A graph $G=(V,E)$.
    }
    \KwOut{A contraction hierarchy $\PCH=(V,\ECH)$}
    \SetKwInOut{Maintains}{Maintains}
      \Maintains{The overlay graph $\overlayG=(\overlayV,\overlayE)$. Initially $\overlayG=G$.\\
  The score of each vertex $\priority[\cdot]$.\\
  }
        $\ECH\gets \emptyset$;
        $\AffectedNeighbors \gets V$; $\WPSSources \gets V$\label{line:init_affected_neighbors}

        \While{$\overlayV \ne \emptyset$}{\label{line:stop_cond}
            \tcp{\textbf{Step 1}: Run witness path searches (WPSes) from all vertices in $\WPSSources$ to prepare for score recomputation for all vertices in $\AffectedNeighbors$}
            \mbox{\upshape \textsf{LocalSearch}$(\WPSSources,\AffectedNeighbors,\overlayG)$}\label{line:pruning} \\
            \tcp{\textbf{Step 2}: For each vertex $v\in \AffectedNeighbors$, (re)compute its score $\priority[v]$}
            \mbox{\upshape \textsf{Score}$(\AffectedNeighbors,\overlayG)$}\label{line:scoring} \\
            \tcp{\textbf{Step 3}: Select an independent set of vertices $\FeasibleCandidates\subseteq \overlayV$ to be contracted based on the score array $P[\cdot]$}
            $\FeasibleCandidates\gets$\mbox{\upshape \textsf{Select}$(\overlayG, P)$}\label{line:selecting}\\
            \tcp{\textbf{Step 4}: Contract vertices in $\FeasibleCandidates$; update $\ECH$ and $\overlayG$ accordingly; return $\prune$ as the sources of WPSes in the next iteration}
            $\WPSSources\gets$\mbox{\upshape \textsf{Contract}$(\FeasibleCandidates,\overlayG)$}\label{line:contracting}\\
            \tcp{$\AffectedNeighbors$ in the next round includes all neighbors of $\FeasibleCandidates$}
            $\AffectedNeighbors\gets\n(\FeasibleCandidates)$\label{line:findneighbor} 
        }

        \Return \mbox{\upshape  \textsf{Postprocess}$(V,\ECH)$}{}\tcp*{reindexing}


\end{algorithm}

In this section, we present \oursys{} for generating contraction hierarchies $\PCH=(V,\ECH)$ efficiently in parallel.
As with other parallel CH algorithms, \oursys{} also follows the high-level idea from Vetter~\cite{vetter2009parallel} that contracts a batch of vertices in an IS in parallel. 
To achieve high parallelism without sacrificing CH quality, \oursys{} uses several novel techniques that will be discussed in this section.

We present the pseudocode of \oursys{} in Alg.~\ref{alg:pch}. 
It takes a graph $G=(V,E)$ as the input, and returns the contracted graph $\PCH$.
Note that the vertex set in $\PCH$ is the same as in the input graph, and the algorithm only needs to compute the edges $\ECH$ of the contracted graph.
Similar to Alg.~\ref{alg:ch}, we also maintain the overlay graph $\overlayG=(\overlayV,\overlayE)$ to reflect the changes in the original graph due to contractions. 

The main loop of our algorithm (the while-loop on line~\ref{line:stop_cond}) repeatedly finds an IS of vertices
and contracts them. We summarize this process in five steps.
We will briefly outline the high-level idea here, and then elaborate on each step.




One of the key insights in \oursys{} is the introduction of the \step{LocalSearch} step on line~\ref{line:pruning}.
This step is used to reduce redundant WPSes, create more parallel tasks, and overall reduce computation while improving parallelism.
With the details of this step described in \cref{sec:prune}, the overall goal of this step is to preprocess the current overlay graph by running WPSes from a set of vertices $\WPSSources$, in order to (re)compute the score for a set of vertices $\AffectedNeighbors$. 
At the beginning, both $\WPSSources$ and $\AffectedNeighbors$ are $V$. 
In later rounds, $\WPSSources$ will be computed by the \step{Contract} step in the previous round, 
and $\AffectedNeighbors$ are the neighbors of vertices contracted in the previous round. 
In Vetter's algorithm, WPSes are directly performed when recomputing scores in each round. 
\oursys{} separates this part with new designs, and further utilizes the results for more optimizations.

The next two steps \step{Score} and \step{Select} will (re-)calculate the score for vertices in $\AffectedNeighbors$, and choose an independent set of vertices $\FeasibleCandidates$ to contract. 
To ensure that selected vertices have low scores, $\FeasibleCandidates$ includes all vertices with the lowest score in their neighborhood.
Importantly, the \step{Score} step makes use of the results from the \step{LocalSearch} step to avoid redundant computation. 

Finally, the \step{Contract} step performs the contraction and updates the overlay graph, 
which is another major improvement over existing solutions. 
In this step, all incident edges to $\FeasibleCandidates$ are moved to $\CH$, and the corresponding shortcuts are added to $\overlayG$.
All these updates in \oursys{} are highly parallel. 
Especially, we use lock-free data structures to update the graph efficiently. 
We introduce the algorithmic idea in \cref{sec:node_contractions}, and discuss the parallel data structure support in \cref{sec:ds}.
In addition, to facilitate the \step{LocalSearch} step, the \step{Contract} step will generate the set of vertices $\WPSSources$, which are all vertices from which we will start WPS in the next round. 

\begin{table}[t]
\small
{\fontsize{8}{9.5}\selectfont\begin{tabular}{@{}lll@{}}
\toprule
$\FeasibleCandidates$:  &  Vertices to be contracted& $\FeasibleCandidates=\{v: \priority[v]<\priority[u]~\forall u\in \n(v)\}$ \\
$\AffectedNeighbors$:  &  Vertices that need to  & $\AffectedNeighbors=\n(\FeasibleCandidates)$ in the previous round\\
 &recalculate their scores& \\
$\ShortcutEdges$: & Shortcuts to be added\\
$\WPSSourcesinit$:  &  Starting points of shortcuts &$\WPSSourcesinit=\{v_1: (v_1,v_2)\in \ShortcutEdges\}$\\
$\WPSSources$:  &  Set of WPS sources & $\WPSSources=\WPSSourcesinit \cup \nin(\WPSSourcesinit)$  \\
\bottomrule
\end{tabular}}
\caption{Additional notations used in our algorithm. \label{tab:our-notation}}
\end{table} 

We summarize the notations in our algorithm in \cref{tab:our-notation}. Next, we will elaborate on each step in Alg.~\ref{alg:pch}.
Later, in \cref{sec:ds}, we discuss using efficient data structures to support this algorithm.

\hide{
\begin{figure}
  \centering
  \includegraphics[width=\columnwidth]{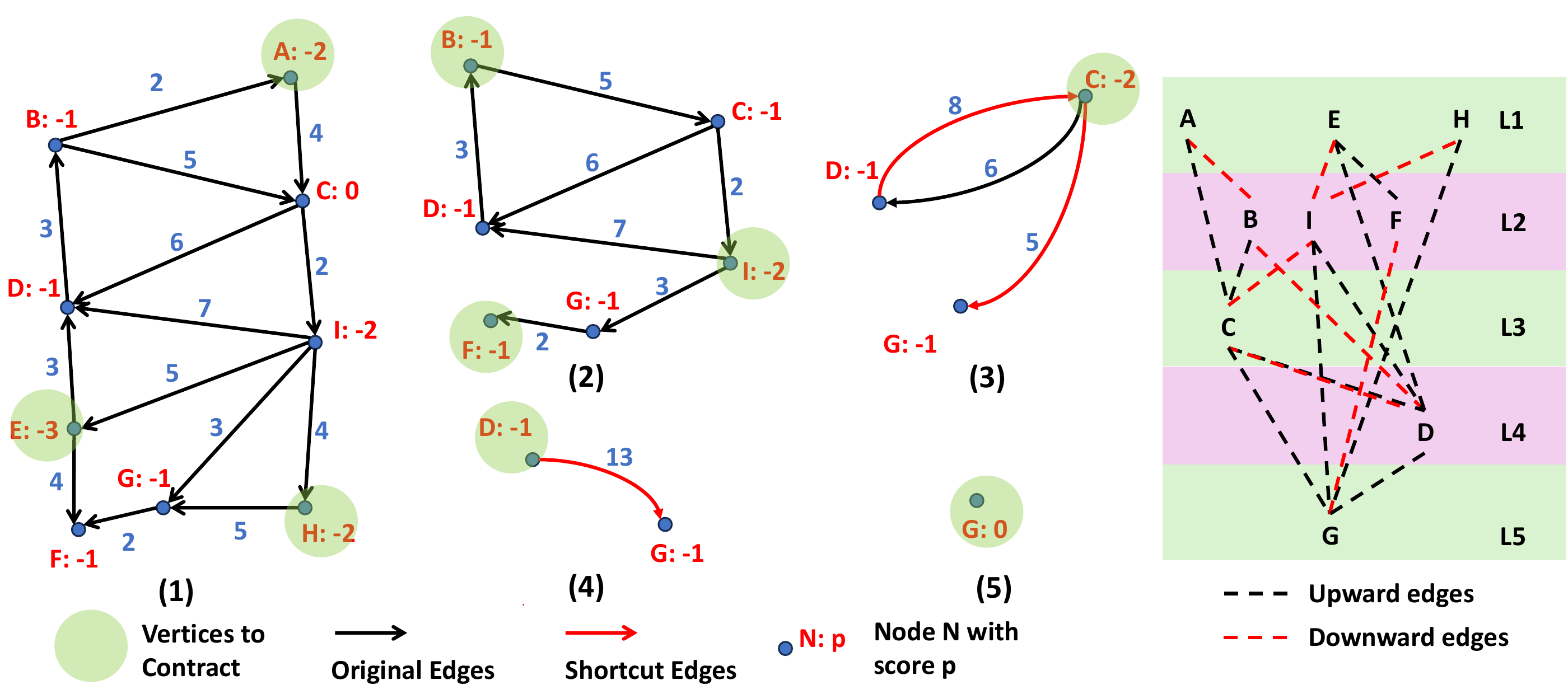}
  \caption{\small \textbf{ An illustration of parallel contraction process}
  }\label{fig:ie_graph}
\end{figure}
\begin{figure}
  \centering
  \includegraphics[width=\columnwidth]{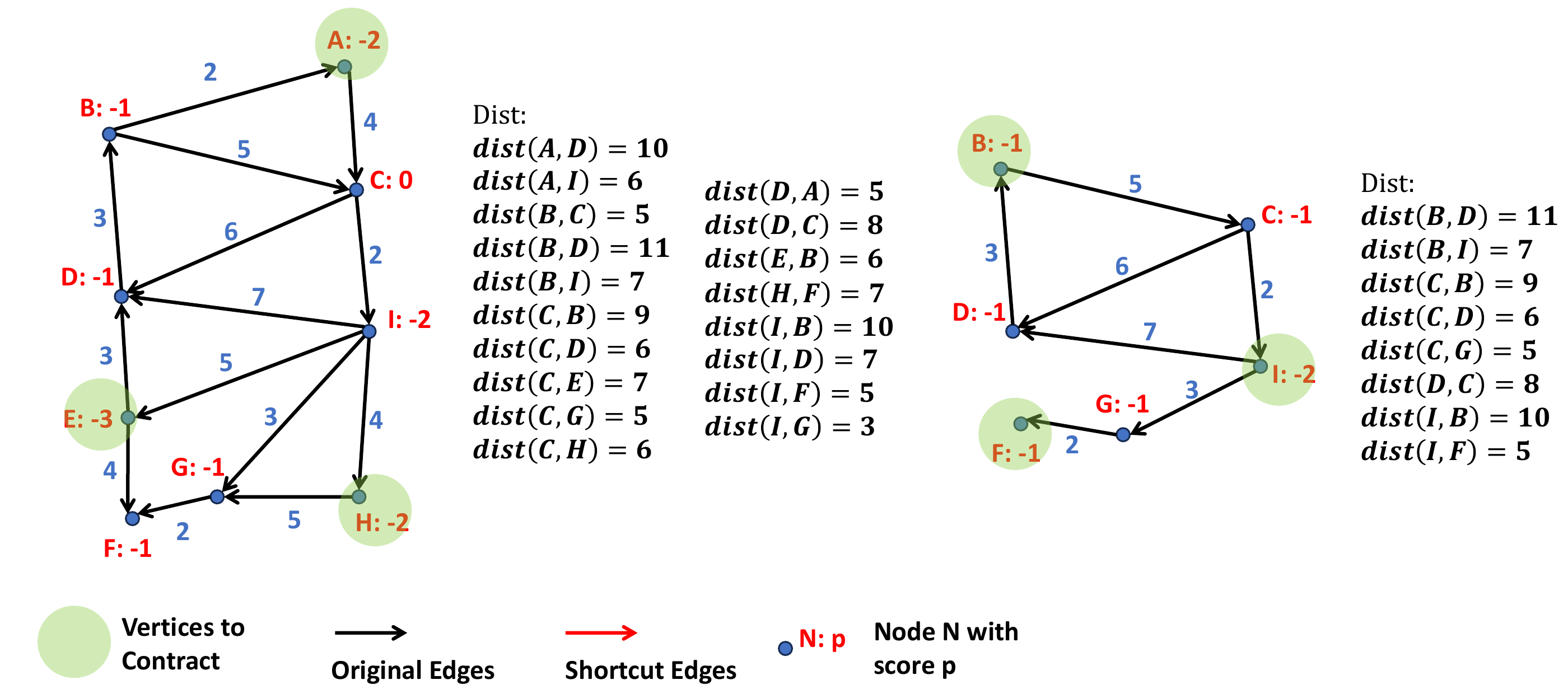}
  \caption{\small \textbf{ Hashmap dist during contraction}
  }\label{fig:per_round_graph}
\end{figure}
}


\subsection{Step 1: Local Search with Memoization}\label{sec:prune}
\begin{algorithm}
    \small
    \caption{Parallel Local Search} \label{alg:prune}
    \DontPrintSemicolon
    \myfunc{\upshape \textsf{LocalSearch}$(\WPSSources,\AffectedNeighbors,\overlayG=(\overlayV,\overlayE))$}{
        \label{line:prune vertices_begin}
        \parForEach{$s$ in $\PruneVertices$}{
            $V_n=\emptyset$\tcp*{the set of vertices that need to be settled by $s$}
            \ForEach {$u\in\nout(s)\land u\in\AffectedNeighbors$}{\label{line:find_new_two_hop_begin}
                \ForEach {$v\in\nout(u): (v\neq s) \land ((s,v)\not\in\dist)$}{
                        $V_n.\batchInsert(v)$\label{line:find_new_two_hop_end}
                }
            }
            \textbf{\boldmath[WPS from $s$]\unboldmath} Run Dijkstra's algorithm starting from $s$. Terminate the algorithm either all $v\in V_n$ are settled, or $\theta$ nodes are settled.
            Add the shortest distance $w$ from $s$ to all $v\in V_n$ into parallel hashmap $\dist$, so that $\dist[(s,v)]=w$\\
            \ForEach{$v$: $(s,v)\in \overlayE$}{
            \lIf{$w(s,v)>\dist[(s,v)]$} {delete edge $(s,v)$ from $\overlayE$}
            }
        }
        \label{line:prune vertices_end}
    }
\end{algorithm} 
One improvement of \oursys{} lies in the stand-alone \step{LocalSearch} step with memoization at the beginning of each round, presented in Alg.~\ref{alg:prune}. 
This step runs witness path searches (WPSes) to prepare the shortest distances needed during scoring (\cref{sec:score}) and contracting (\cref{sec:node_contractions}).
As mentioned in~\cref{sec:och-construct}, to compute the score of a vertex $v\in \AffectedNeighbors$, 
a simulated contraction on $v$ is needed to determine the shortcuts that would be added if we contract $v$.
For each pair of vertices $(u_1,u_2)$ where $u_1\in \nin(v)$ and $u_2\in \nout(v)$,
if $v$ is on the shortest paths from $u_1$ to $u_2$, then a shortcut is needed. 
We also use the WPS as described in \cref{sec:och-construct},
which runs Dijkstra from $u_1$ until $u_2$ is settled to find the shortest distance between them. 

Running WPSes is crucial to determine the score of a vertex. 
However, it is also expensive---to compute the edge difference for a vertex $v$, 
WPSes are needed on all pairs in $\nin(v) \times \nout(v)$ (Cartesian product).  
Fully finishing WPSes for all pairs may be expensive, especially when there exists one pair of vertices that are extremely far away. 
Therefore, most existing systems (including ours) also limit the size of each WPS to be $\theta$, i.e., each Dijkstra's algorithm only searches for $\theta$ iterations. In this case, some suboptimal shortcuts may be added and make the CH (necessarily) larger. 
To improve the overall performance, we highlight three key techniques in our \step{LocalSearch} step: batching, memoization, and pruning. 

\begin{figure}[t]
    \centering
    \includegraphics[width=\columnwidth]{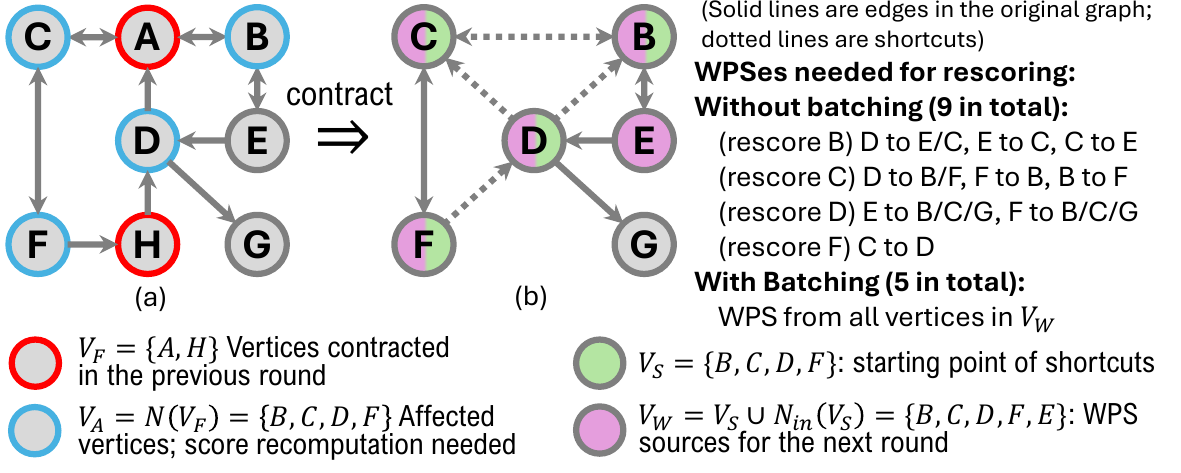}
    \caption{\small \textbf{An illustration of notations and benefit of batching in \oursys{}.}
    The two figures illustrate the vertex sets $\AffectedNeighbors, \FeasibleCandidates,\WPSSourcesinit$, and $\WPSSources$. A node in both purple and green means it is in both $\WPSSources$ and $\WPSSourcesinit$. 
    On the right, we show an illustration of using batching to save WPSes. To rescore the four vertices in $\AffectedNeighbors$, 
    for each $v\in \AffectedNeighbors$, previous solutions will perform WPSes on all pairs $\nin(v)\times \nout(v)$. 
    \oursys{} identifies all possible WPS sources in the \step{Contract} step in the previous round and collect them in $\WPSSources$.
    Therefore, the WPSes from the same source will be conducted only once. In this example, \oursys{} only need 5 WPSes instead of 9. 
      }\label{fig:local-search-benefit}
\end{figure}

\myparagraph{Batching.} Our first technique is to collect all WPSes needed and process them in a batch. 
In \oursys{}, the \step{Contract} step generates a vertex set $\WPSSources$, 
which serves as sources of WPSes in the next round. 
By maintaining them in a batch, we process all WPSes in a batch to avoid redundant computation and achieve better parallelism. 
For example, if two simulated contractions trigger WPSes from the same vertex, they will be performed only once in the batch.
An illustration of all WPS sources and how batching saves the number of WPSes is given in \cref{fig:local-search-benefit}. 
Executing all WPSes as one batch allows us to optimize them as a whole, and provides greater opportunity to exploit parallelism. 

\myparagraph{Memoization.} \oursys{} uses memoization to avoid computing the shortest distance on the same vertex pair multiple times.
In \oursys{}, a map $\dist$ is used to memoize pairwise shortest distances computed in previous WPSes.
If a WPS is required on the same pair again, we directly use the memoized results without running another WPS. 
This information is reused in pruning (described below) and later steps \step{Score} and \step{Contract}.

\myparagraph{Pruning.} A byproduct of memoizing the distances in $\dist$ is to prune redundant edges in $\overlayG$. 
As the algorithm proceeds, more pairs of vertices $(u,v)$ obtain their true shortest distance $d(u,v)$ by WPSes (using Dijkstra's algorithm). 
If there is an edge between $u$ and $v$ in the current overlay graph (possibly a suboptimal shortcut added in previous rounds) with weight $w>d(u,v)$, the edge can be deleted since it is redundant.
This optimization effectively reduces the size of the overlay graph, which potentially improves the performance for both construction and query.

\myparagraph{The \step{LocalSearch} Step.} With the three techniques, we now describe the \step{LocalSearch} step. 
The step takes $\WPSSources$, $\AffectedNeighbors$ and $\overlayG$ as input, and performs WPSes from all vertices in $\WPSSources$ to recompute the scores for all vertices in $\AffectedNeighbors$. 
For each WPS from $s$, we first identify the set of vertices $V_n$ that need to be settled by $s$.
On line~\ref{line:find_new_two_hop_begin}--\ref{line:find_new_two_hop_end} in Alg.~\ref{alg:prune}, for each $u\in\nout(s)$, if $u\in\AffectedNeighbors$,
this means that $s$ is a source of WPS because $s$ is an in-neighbor of an affected vertex $u$.
To compute the new score for $u$,
the shortest distances from $s$ to each vertex $v\in \nout(u)$ need to be computed.
Therefore, we collect $\nout(u)$, the out-neighbors of $u$, to the set $V_n$.
Note that $s$ can correspond to multiple affected vertices $u$.
In this case, all of their out-neighbors will be gathered in $V_n$, and only incur one search starting from $s$.
Before adding a vertex $v$ to $V_n$, we also check if $(s,v)$ has already been memoized in $\dist$. If so, $v$ will be skipped.

With all target vertices of WPS originating from $s$ gathered in $V_n$, 
We run Dijkstra's algorithm from $s$ to search for all vertices in $V_n$.
During the algorithm, the shortest distances from $s$ to each $v\in V_n$ are inserted into $\dist$. 
As mentioned, a WPS terminates either when all vertices in $V_n$ have been settled, or when $\theta$ vertices have been settled (so that the overall cost is limited). 
If the WPS terminates before a vertex $v\in V_n$ is settled, the distance between $s$ and $v$ will be viewed as $\infty$ and a shortcut between $s$ and $v$ is always generated.

\subsection{Step 2: Score}\label{sec:score}
\begin{algorithm}
\DontPrintSemicolon
    \small
    \caption{Parallel Node Scoring} \label{alg:score}
    \label{line:score_begin}
    \myfunc{\upshape \textsf{Score}$(\AffectedNeighbors,\overlayG=(\overlayV,\overlayE))$}{
        \parForEach{$u$ in $\AffectedNeighbors$}{
            $\priority[u]\gets -|\n(u)|$ \label{line:init_priority}
            \\\ForEach {$v_1\in\nin(u)$}{
                \ForEach {$v_2\in\nout(u)$}{
                    \lIf {$w(v_1,u)+w(u,v_2) = \dist\langle v_1,v_2\rangle \lor (v_1,v_2)\not\in\dist$}{
                        $\priority[u] \gets \priority[u] + 1$
                    }
                }
            }
        }
    }
    \label{line:score_end}
\end{algorithm} 
The second step computes the scores for each vertex $u\in\AffectedNeighbors$, as presented in Alg.~\ref{alg:score}. 
In the previous step, we have memoized the distances in the neighborhood of each $u\in\AffectedNeighbors$ in a map $dist$.
To recompute the score of $u\in \AffectedNeighbors$, we first set $\priority[u]$ to the negative of its number of direct neighbors, i.e., $-|\n(u)|$, since $|\n(u)|$ edges can be removed from $\overlayG$ if $u$ gets contracted (line~\ref{line:init_priority}).
Then, we iterate through each pair of in-neighbors and out-neighbors of $u$ and check if their distance through $u$ is equal to or smaller than the shortest distance stored in $\dist$. If so, we add one to $\priority[u]$ since one shortcut needs to be added.

For simplicity, our pseudocode only computes the edge difference, which is the most important component of the score for each vertex.
In practice, many other components are considered in existing work. 
\oursys{} uses a similar score function as in previous work, which is a combination of edge difference, vertex degree, hop distance, and hierarchy depth. 

\subsection{Step 3: Select}\label{sec:select}
\begin{algorithm}
    \small
    \DontPrintSemicolon
    \caption{Parallel Node Selection} \label{alg:select}
    \myfunc{\upshape \textsf{Select}$(\overlayG=(\overlayV,\overlayE))$}{
        $\FeasibleCandidates\gets \emptyset$\\
        
        \parForEach{$u$ in $\overlayV$}{\label{line:select_begin}
            \lIf{$(\forall v\in\n(u), \priority[u]<\priority[v])$}{
                $\FeasibleCandidates.\text{insert}(u)$
            }
        }
        \label{line:select_end}
        \Return{$\FeasibleCandidates$}
    }
\end{algorithm} 
The \step{Select} step identifies a subset of vertices $\FeasibleCandidates$ to contract in parallel, as presented in Alg.~\ref{alg:select}.
We aim to contract multiple vertices in an independent set (i.e., do not share edges) simultaneously while prioritizing those with lower scores.
To do this, we select all vertices that have the minimum score in its neighborhood.
As shown in line~\ref{line:select_begin}--\ref{line:select_end} in Alg.~\ref{alg:select}, we process each vertex $u$ in $\overlayV$.
If $u$ has the lowest score among its neighbors, we insert $u$ into a set $\FeasibleCandidates$.
To handle equal scores, we use the label of each vertex to break ties.
This approach can find an independent set so that vertices with lower scores are contracted before those with higher scores.
\cref{fig:ch-example}(b) presents an example of contracting a graph in parallel. In each round, multiple independent vertices can be contracted together,
and the levels can be determined accordingly. The final graph preserves the shortest distance between any two vertices.

Selecting an IS of vertices ensures that shortcuts are always established between non-contracted vertices.
The process is also reasonably fast for contraction. 
In our experiments, we observe that usually within 10--20 rounds, more than 99\% of vertices are contracted, indicating high potential of parallelism in this algorithm.  

\subsection{Step 4: Contract}\label{sec:node_contractions}
\begin{algorithm}
    \small
    \caption{Parallel Node Contraction} \label{alg:contract}
    \myfunc{\upshape \textsf{Contract}$(\FeasibleCandidates,\overlayG=(\overlayV,\overlayE))$}{
        $\WPSSourcesinit\gets\emptyset$\\
        \parForEach{$u$ in $\FeasibleCandidates$}{
            Remove $u$ from $\overlayV$\label{line:move_vertex}
            \\
            \lForEach {$v\in\nin(u)$}{
                $\BackwardEdges.\batchInsert(\langle v,u,w(v,u) \rangle)$
            }
            \label{line:move_edges_begin}
            \lForEach {$v\in\nout(u)$}{
                $\ForwardEdges.\batchInsert(\langle u,v, w(u,v) \rangle)$
            }
            \label{line:move_edges_end}

            \For {$v_1\in\nin(u)$\label{line:add_shortcut_begin}}{
                \For {$v_2\in\nout(u)$}{
                    \If {$w(v_1,u)+w(u,v_2) = \dist[(v_1,v_2)] \lor (v_1,v_2)\not\in\dist$\label{line:contract:if}}{
                        $\WPSSourcesinit.\batchInsert(v_1)$\\
                        $\ShortcutEdges.\batchInsert(\langle v_1,v_2, w(v_1,u)+w(u,v_2) \rangle)\label{line:contract:add_shortcut}$
                    }
                }
            }
            \label{line:add_shortcut_end}
            Remove the incident edges of $u$ from $\overlayE$\label{line:removeEO}
        }

        $\overlayE=\overlayE\bigcup\ShortcutEdges$ \label{line:contract:combine} \\
        $\WPSSources\gets\emptyset$\\
        \parForEach{$u$ in $\WPSSourcesinit$}{
            $\WPSSources.\text{insert}(u)$
            \\\lparForEach{$v\in\nin(u)$}{
                $\WPSSources.\text{insert}(v)$
             }
        }
        \Return{$\WPSSources$}
    }
\end{algorithm} 
With the independent set $\FeasibleCandidates$ computed in the previous round,
the last step \step{Contract} performs the actual contraction, as given in Alg.~\ref{alg:contract}.
This step moves all vertices in $\FeasibleCandidates$ from $\overlayV$ to the CH, adds incident edges to $\ECH$,
computes and adds the relevant shortcuts to $\overlayE$, and finally generates the set $\WPSSources$ as the sources for running WPSes in the next round,
which will be used in the next \step{LocalSearch} step.

In Alg.~\ref{alg:contract}, we process each vertex $u\in \FeasibleCandidates$ in parallel.
We first remove each vertex $u\in\FeasibleCandidates$ from the overlay vertex set $\overlayV$ in line~\ref{line:move_vertex}.
Next, as shown in line~\ref{line:move_edges_begin}--\ref{line:move_edges_end} in Alg.~\ref{alg:contract}, we move the incident edges of $u$ to $\residualE$, which contains two subsets: $\BackwardEdges$ and $\ForwardEdges$.
When moving a vertex $u$ to the CH, its incoming edges are stored in $\BackwardEdges$, and outgoing edges are stored in $\ForwardEdges$.
This separation is necessary because during the shortest path queries, the search from the target only proceeds ``backward'' by following incoming edges in $\BackwardEdges$, while the search from the source only moves ``forward'' by following outgoing edges in $\ForwardEdges$. Both searches move ``upward'' in the CH.
\hide{
This separation is necessary because during the shortest path queries, the search from the target only proceeds ``downwards'' by following incoming edges in $\BackwardEdges$ while the search from the source only moves ``upwards'' by following outgoing edges in $\ForwardEdges$.
}

Finally, as shown in line~\ref{line:add_shortcut_begin}--\ref{line:add_shortcut_end} in Alg.~\ref{alg:contract}, for each vertex $u$ to be contracted,
we enumerate each in-neighbor $v_1\in \nin(u)$ and out-neighbor $v_2\in \nout(u)$.
We establish a shortcut between $v_1$ and $v_2$ if necessary.
This is performed by comparing $w(v_1,u)+w(u,v_2)$ with the shortest distance stored in $\dist$.
Recall that $\dist$ is a map generated in the \step{LocalSearch} step,
which buffers shortest distances for relevant vertex pairs.
If $w(v_1,u)+w(u,v_2)=\dist[(v_1,v_2)]$, then the shortest path between $v_1$ and $v_2$ can be via $u$.
A special case is when $(v_1,v_2)$ is not in $\dist$, which means that the shortest distance between them was not computed in the \step{LocalSearch} step.
In both cases, before contracting $u$, a shortcut needs to be established between $v_1$ and $v_2$ with weight $w(v_1,u)+w(u,v_2)$ (line~\ref{line:contract:if}).

\begin{figure}[t]
    \centering
    \includegraphics[width=\columnwidth]{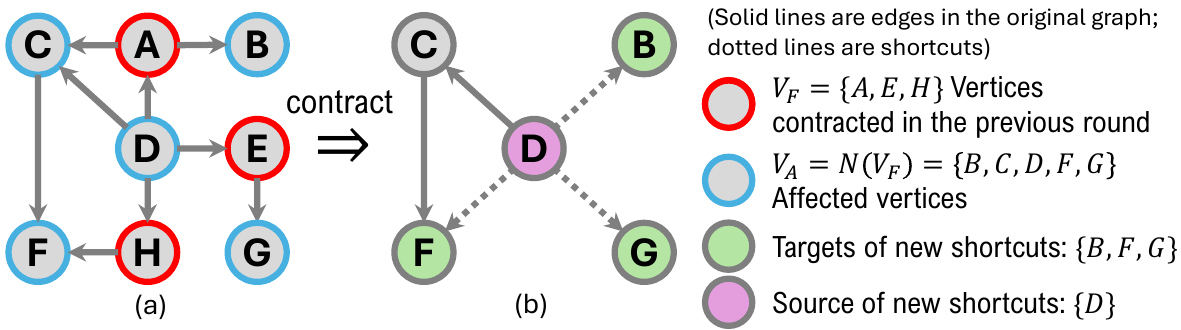}
    \caption{\small\textbf{An illustration of concurrent shortcut insertions in \oursys{}.}
    During the \step{Contract} step, for every feasible candidate \(u \in \FeasibleCandidates\), a dedicated thread inserts shortcuts between each pair \((v_1,v_2)\) with \(v_1 \in \nin(u)\) and \(v_2 \in \nout(u)\).
    In the example, three threads simultaneously attempt to insert edges $(D,B)$, $(D,F)$, $(D,G)$, which share the same source $D$, potentially creating a race condition. }
    \label{fig:shortcut-insertion}
\end{figure}
However, a write–write race condition may occur when multiple shortcuts that share the same source or target node are inserted in parallel.
An example of this issue is shown in~\cref{fig:shortcut-insertion}.
Vertex $D$ has three out-neighbors that are being contracted, each introduces a new out-neighbor to $D$.
Since the contractions of these three original out-neighbors, $\{A, E, H\}$, run concurrently, the insertions of $D$'s new out-neighbors, $\{B, F, G\}$, also happen in parallel.
Therefore, we need a parallel data structure that supports frequent and concurrent insertions, while providing fast query performance.
In our implementation, we use the parallel hash map $\ShortcutEdges$ in \cref{sec:data-structure-background} to buffer all shortcuts added in this round and later combine it with $\overlayE$ on line~\ref{line:contract:combine}. In \cref{sec:use-data-structure} we will describe other optimizations to reduce the cost to combine $\ShortcutEdges$ and $\overlayE$.
Finally, we can also remove all edges incident on $u \in \FeasibleCandidates$ from $\overlayE$ after adding shortcuts.
This is because these edges no longer contribute to the reduction of distances to any vertices that remain uncontracted.


\hide{For example, to contract vertex $u$ with incident edges $\langle v_1,u, w_1 \rangle$ and $\langle u,v_2,w_2\rangle$.
A shortcut from $v_1$ to $v_2$ is needed when the shortest paths from $v_1$ to $v_2$ is through $u$.
As mentioned before in~\cref{sec:prune}, in order to verify this, witness path searches on all the new shortcuts have been conducted and the shortest distances for all relevant $(v_1,v_2)$ pairs have been buffered in $\dist$.
If the shortest distance between $v_1$ and $v_2$ stored in $\dist$ equals to $w_1+w_2$,
a shortcut is needed from $v_1$ to $v_2$.
In this case, we move $\langle v_1, u, w_1 \rangle$ to $\BackwardEdges$, $\langle u,v_2,w_2\rangle$ to $\ForwardEdges$, and add shortcuts $\langle v_1, v_2, w_1+w_2 \rangle$ to $\overlayE$.
In our implementation, we use a parallel hash map $\ShortcutEdges$ in \cref{sec:impl-details} to buffer all shortcuts added in this round and later combine it with $\overlayE$. In \cref{sec:xx} we will describe other optimizations to reduce the cost to combine $\ShortcutEdges$ and $\overlayE$.
Finally, we can also remove all edges incident on $u \in \FeasibleCandidates$ from $\overlayE$ after adding shortcuts.
This is because these edges no longer contributes to the reduction of distances to any vertices that remain uncontracted.
}

During the contraction step, we also generate the set $\WPSSources$ that contains all sources for the WPSes in the next step.
When contracting a vertex $u$, we check all $v_1\in \nin(u)$ and $v_2\in\nout(u)$ to see whether a shortcut from $v_1$ to $v_2$ needs to be established.
If so, the scores for both $v_1$ and $v_2$ are affected, since both have their neighborhoods changed.
To reconsider the score of $v_2$, WPSes are needed from $v_1$ (to find the distance from the new in-neighbor $v_1$ to all $\nout(v_2)$).
To reconsider the score of $v_1$, WPSes are needed from all in-neighbors of $v_1$ (to find their distances to the new out-neighbor $v_2$).
Hence, we first use a set $\WPSSourcesinit$ to buffer all vertices $v_1$.
At the end of this step, we construct the set $\WPSSources$, which includes all vertices in $\WPSSourcesinit$ and all in-neighbors of them.
This set will be used as the sources for the WPSes in the \step{LocalSearch} step in the next round.
An illustration of the vertex sets $\WPSSourcesinit$ and $\WPSSources$ is shown in \cref{fig:local-search-benefit}.

\subsection{Postprocessing}
After contracting all vertices, we re-index and sort the vertices~\cite{blelloch2010low,dong2024parallel} by their levels and scores to improve query performance.
In the sequential CH algorithm~\cite{geisberger2008contraction}, vertices are reordered by the order in which they are contracted. 
In \phast~\cite{PHAST}, vertices are grouped by levels to perform SSSP queries on the CH, where lower level vertices have a lower rank than those in higher levels.
We combine the two re-indexing methods and use vertex scores to approximate the contraction order.
For any two vertices, if they are at the same level, they are sorted by score from the lowest to the highest.
Otherwise, they are sorted by level from the lowest to the highest.

This re-indexing ensures that the vertices are efficiently organized for query processing, where paths are traversed based on the contraction rank.
Once the vertices are re-indexed, we generate the final CH from $\ForwardEdges$ and $\BackwardEdges$. 

\section{Parallel Data Structures}\label{sec:ds}


In \cref{sec:contraction}, we presented the algorithmic idea of \oursys{}.
We note that, however, to support high performance in practice,
the implementation relies on efficient data structures to support the algorithm, especially in handling dynamic updates of the overlay graph.  
Note that the (overlay) graph is highly dynamic---as the contraction proceeds,
the graph consistently undergoes vertex removals (removing contracted vertices),
edge removals (removing edges incident on contracted vertices), and edge insertions (adding shortcuts).
Each of these operations require high parallelism to enable good performance for the overall CH algorithm.
This section provides an overview of the parallel data structures used in our implementation.

Especially, to facilitate graph traversal, the overlay graph is usually maintained by an array-based structure (e.g., the CSR introduced below),
which maintains the neighbors of the same vertex in a consecutive chunk in the memory.
When multiple vertices are contracted in parallel, shortcuts are also generated by multiple threads.
Inserting them into the overlay graph in a concurrent manner is therefore challenging.
In existing implementation, \osrm{} avoids such complications by only contracting vertices with the minimum score in its 2-hop neighborhood.
In this case, for a vertex $v$ in the overlay graph, only one contracted vertex may update the neighbor list of $v$, and this will be performed sequentially.
However, by restricting all contracted vertices to be at least two hops away, \osrm{} can choose far fewer vertices in the independent set in each round.
In our experiments, we observe that such restriction greatly increase the number of rounds and degraded the performance.
In \chconstructor{}, a lock-based data structure is used, which also limits parallelism.
In this section, we introduce the parallel data structures used in \oursys{}, such that they support updating the overlay graph in an efficient and flexible way.

\newcommand{\mf}[1]{{\mbox{\sc{#1}}}}
\newcommand{\bagput}{\mf{BagInsert}\xspace}
\newcommand{\bagpack}{\mf{BagExtractAll}\xspace}

\subsection{Background of Existing Data Structures}\label{sec:data-structure-background}

\oursys{} mainly uses two data structures to maintain graphs: Compressed Sparse Row (CSR)~\cite{tinney1967direct} and phase-concurrent hash table~\cite{shun2014phase}.
In this section, we introduce the background of these data structures and their interface.
Then in \cref{sec:use-data-structure}, we describe how they are used to support our algorithm.


\myparagraph{Compressed Sparse Row (CSR)~\cite{tinney1967direct}.}
CSR is a widely used array-based adjacency list representation in graph algorithms~\cite{dong2021efficient,dong2023provably,liu2025parallel,wang2023parallel,dhulipala2024optimal,dong2024pasgal,gbbs2021,wheatman2024byo},
consisting of an \texttt{edge} array of size $|E|$
and an \texttt{offset} array of size $|V|$.
The \texttt{edge} array is a concatenation of neighbor lists (along with the corresponding edge weights) for all vertices.
The neighbor list for each vertex is always contiguous.
The \texttt{offset} array stores the start position of each vertex's neighborhood in the \texttt{edge} array.
CSR provides \emph{fast lookups with good cache locality} due to its contiguous memory layout.
However, it does not support insertion or deletion without rebuilding the entire edge and offset arrays.
Our input graph is given in the CSR format.

\myparagraph{Phase-Concurrent Hash Table~\cite{shun2014phase}.}
The phase-concurrent hash table is an array-based structure that supports efficient operations for an unordered set,
including insertions, deletions, and lookups. We use it to support multiple variables in \oursys{}.

Unlike traditional concurrent hash tables that allow arbitrary operations of any type to run concurrently,
the phase-concurrent hash table only permits operations of the same type to proceed simultaneously.
This design ensures both \emph{serialization} and \emph{determinism} (w.r.t. execution order).
To insert a key-value pair $(k,v)$ into the hash table $H$,
the key $k$ is first hashed to a random index in $H$.
If this index is occupied, linear probing is performed to find the next available slot.
Once an empty index is found, it swaps in the new record using an atomic operation \cas{}.
When multiple threads attempt to modify the same memory location, the atomic operation guarantees that only one thread succeeds.
All other threads will fail in the operation and continue linear probing.
A special case for insertion is to update the value for the $k$ if it already exists in the hash table.
In our implementation where the value can refer to the weight of a given edge,
we update the value if the newly inserted value is smaller than the current value stored in the hash table.
In particular, when inserting $(k,v)$, if we find that $k$ is already in the table during linear probing,
we use an atomic operation \WriteMin{} to update the value and keep the lower one.
The atomicity of this operation guarantees that only the minimum value among all concurrent modifications is retained.

Similarly, to look up a record with key $k$,
the hash table hashes $k$ to an initial index $i$.
Then, three cases are considered:
1) If $H[i]$ is empty, then $k$ does not exist in the hash table.
2) If the key at $H[i]$ is $k$, then the record with key $k$ has been found.
3) If the key at $H[i]$ is not $k$, the algorithm uses linear probing to continue searching for the key in subsequent indices, until $k$ is found or an empty slot is encountered.
To delete a record with key $k$, the algorithm first locates $k$ in $H$.
If $k$ is present, the corresponding index is marked as a tombstone,
which is not treated as empty but will not match any existing keys.
If $k$ is not found in $H$, no further action is needed.

For simplicity, we will refer to the phase-concurrent hash table as the \emph{hash table} in the following part of this section.

\subsection{Using Parallel Data Structures in \oursys{}} \label{sec:use-data-structure}

\myparagraph{Maintaining the Overlay Graph Dynamically.}
To facilitate graph traversal, the overlay graph $\overlayG=(\overlayV,\overlayE)$ is maintained using a CSR, such that getting the neighbor list for each vertex is always efficient.
The challenge is then to efficiently reflect edge insertions in the static CSR structure.

To efficiently support these updates in the overlay graph, we use an auxiliary edge set $\ShortcutEdges$ to tentatively maintain the newly added edges.
Recall that when a node $u$ is contracted, pairwise edges from $v_1\in \nin(u)$ to $v_2\in \nout(u)$ need to be added to the overlay graph if the shortest path from $v_1$ to $v_2$ passes through $u$.
Instead of adding them directly to the CSR of the overlay graph, we first use $\ShortcutEdges$ to buffer them.
Once a shortcut is generated (line~\ref{line:contract:add_shortcut} in Alg.~\ref{alg:contract}), we directly add the new edge to $\ShortcutEdges$.
However, adding these edges to $\ShortcutEdges$ in parallel is challenging due to contention---although contracting two adjacent vertices is avoided by selecting an IS,
it is possible for multiple vertices to be contracted share the same neighbor.
In this case, new edges will be concurrently added to this neighbor.
Moreover, edges with the same endpoints $(u,v)$ could be added to the overlay graph by two different contracted vertices $c_1$ and $c_2$ (i.e., via $u$--$c_1$--$v$ and $u$--$c_2$--$v$).
Among all the same edges, only the one with the smallest weight should be kept.

To support these updates in the overlay graph, we use a parallel hash table to maintain $\overlayE$.
Once a shortcut $(v_1,v_2)$ with weight $w$ is generated, we insert the triple $(v_1,v_2,w)$ to the hash table.
To insert an edge $(v_1,v_2,w)$,
we first check whether the edge $(v_1,v_2)$ exists in the hash table.
If so, we use \WriteMin{} to update the weight atomically if $w$ is smaller than the stored weight.
Otherwise, we insert $(v_1,v_2,w)$ to the hash table.
Here we use $v_1$ as the key, and we will explain the design choice later in this section.

Note that even if we first collect shortcuts in $\ShortcutEdges$ and combine them with $\overlayE$ to avoid frequently updating the CSR,
it can still be expensive to do so every round.
Especially in later rounds, only a few vertices are contracted, generating a small number of shortcuts.
In this case, updating the CSR to add a few edges is inefficient.
To handle this issue, we combine the two sets in a lazy manner.
In particular, we do not clear and combine $\ShortcutEdges$ with $\overlayE$ until $\ShortcutEdges$ gets overloaded (i.e., a constant fraction of slots are taken).
We set the hash table size to the number of edges in the initial CSR.
This guarantees that when updating the CSR, there must be a comparable number of edge insertions, and thus the cost can be amortized.

However, the belated combination of $\ShortcutEdges$ and $\overlayE$ also requires all edges in $\ShortcutEdges$ to be visible to later graph traversal operations.
To do this, when inserting an edge $(v_1,v_2,w)$, we use $v_1$ as the key and $(v_2,w)$ as the value.
Using only the first endpoint as the key allows us to look up the neighbors of a given vertex by linear probing, and also has better cache locality.
When traversing all neighbors of $u$ in the overlay graph, we first process all neighbors of $u$ in the CSR $\overlayE$.
We then further look up $u$ in the hash table for $\ShortcutEdges$, and use linear probing to get all edges incident on $u$.
These provide the additional neighbors of $u$ w.r.t. shortcuts associated with $u$ that have not yet been incorporated into $\overlayE$.

To identify when the hash table is overloaded, at the end of each round, we use random sampling to estimate the size of the hash table.
If the hash table is filled up by a constant fraction, a merge of $\ShortcutEdges$ with $\overlayE$ is triggered.
During this process, we also remove all edges incident to the vertices that have been contracted.
In other words, the edge removals in line~\ref{line:removeEO} of Alg.~\ref{alg:contract} are also performed lazily
until the next update on $\overlayE$.
Finally, we clear the hash table.


The final edges in the CH, $\ForwardEdges$ and $\BackwardEdges$ are also maintained by the phase-concurrent hash tables.
For both of them, only the insertion operation is needed, which can be performed concurrently using atomic operations.

\myparagraph{Maintaining the Distance Mapping $\dist$.} Another important data structure used in our algorithm is $\dist$, which memoizes
shortest distances for certain pairs of vertices discovered in WPSes.
In our implementation, we also use a phase-concurrent hash table.
When a shortest distance between $u$ and $v$ is computed as $w$, we add $(u,v,w)$ to the hash table, where $(u,v)$ is the key and $w$ is the value.
By setting up the mapping in the \step{LocalSearch} step, later \step{Score} and \step{Contract} steps can directly look up the distance instead of performing another simulated contraction.

Similar to the shortcut set $\ShortcutEdges$, the number of elements in the $\dist$ map can also change dramatically in different rounds.
When most slots in the hash table are empty, clearing the map becomes expensive. To address this issue, similar to
how we maintain $\ShortcutEdges$, we clear the hash table for $\dist$ lazily. At the end of the round, we use sampling to estimate
the load factor of the hash table.
Only when a constant fraction of the hash table is full do we clear the hash table.

\section{Query on Contraction Hierarchies}\label{app:query}
In this section, we describe the standard query algorithm on the CH.
After constructing the CH, we re-index and sort all vertices by their level and score.
In Geisberger's sequential CH algorithm~\cite{geisberger2008contraction}, vertices are reordered based on the order to contract them.
$\rank(v)$ is the order to contract $v$.
In \PHAST~\cite{delling2013phast}, vertices are re-grouped by levels to perform SSSP queries on CH, where vertices in lower levels always have a lower rank than those in higher levels.
In the query phase, all the edges in $\ECH$ are divided into two parts and forming two graphs:
\hide{
$\ForwardEdges$, which contains the edges leading to neighbors with higher rank, and $\BackwardEdges$, which contains the edges leading to neighbors with lower rank.
}
the out-edges from $u$ to higher-ranked vertices are referred to as the \defn{upward edges}:
$\ForwardEdges \gets \{(u,v)~|~\rank[u] < \rank[v]\}$,
while the in-edges from higher-ranked vertices to $u$ as the \defn{downward edges}:
$\BackwardEdges \gets \{(v,u)~|~\rank[u] < \rank[v]\}$.
For each edge connecting two vertices, only the endpoint of the lower rank stores this edge in $\CH$.

CH boost up the efficiency of point to point query ($s$-$t$ query) significantly.
During a point-to-point query from $s$ to $t$, a forward search from $s$ is performed on $\ForwardEdges$ and a backward search from $t$ is performed on $\BackwardEdges$, meaning that only edges leading to higher rank vertices are considered.
For each vertex $v$, the distance from $s$ to $v$ and the distance from $v$ to $t$ is maintained as $d_s(v)$ and $d_t(v)$ and the estimated distance from $s$ to $t$ through $v$ is $\dist_s(v)+\dist_t(v)$.
Once the tentative shortest distance $\mu$ from $s$ to $v$ is not larger than the minimum value of the priority queue, the search result is settled to $\mu$.
\section{Experiments}\label{sec:exp}

\aptLtoX[graphic=no,type=html]{\begin{table*}
  \small
  \centering
  \setlength{\tabcolsep}{3pt}
    \begin{tabular}{cl|c@{\,\,\,}c@{\,\,\,}c@{\,\,\,}c@{\,\,\,}c@{\,\,\,}c|c@{\,\,\,}c@{\,\,\,}c@{\,\,\,}c|c@{\,\,\,}c@{\,\,\,}c@{\,\,\,}c@{\,\,\,}c@{\,\,\,}c}
    \multicolumn{2}{c|}{\multirow{2}[0]{*}{\textbf{Graph}}} & \multicolumn{6}{c|}{\textbf{Road}}            & \multicolumn{4}{c|}{\textbf{Synthetic}} & \multicolumn{6}{c}{\textbf{$k$-NN}} \\
    \multicolumn{2}{c|}{} & \textbf{CA} & \textbf{AO} & \textbf{AF} & \textbf{NA} & \textbf{AS} & \textbf{EU} & \textbf{TRCE} & \textbf{BUB} & \textbf{CHN7} & \textbf{CHN8} & \textbf{HT5} & \textbf{CH2} & \textbf{CH5} & \textbf{GL2} & \textbf{GL5} & \textbf{GL10} \\
    \multicolumn{2}{c|}{\textbf{\# vertices}} & 3.34M & 6.22M & 33.5M & 87.0M & 95.7M & 131M  & 16.0M & 21.2M & 10.0M & 100M  & 929K  & 4.21M & 4.21M & 24.9M & 24.9M & 24.9M \\
    \multicolumn{2}{c|}{\textbf{\# edges}} & 4.20M & 8.24M  & 44.8M & 113M  & 123M  & 169M  & 48.0M & 63.6M & 10.0M & 100M  & 4.64M    & 8.42M & 21.0M & 49.8M & 124M  & 249M \\
    \midrule
    \multicolumn{1}{r}{\multirow{8}[2]{*}{\textbf{Build (s)}}} &  \textbf{\oursysTab*} & \underline{\textbf{1.00}} & \underline{\textbf{2.25}} & \underline{\textbf{7.32}} & \underline{\textbf{23.1}} & \underline{\textbf{23.5}} & \underline{\textbf{32.9}} & \underline{\textbf{25.2}} & \underline{\textbf{34.7}} & \underline{\textbf{1.77}} & \underline{\textbf{14.9}} & \underline{\textbf{0.68}} & \underline{\textbf{0.87}} & \underline{\textbf{115}} & \underline{\textbf{2.95}} & \underline{\textbf{9.78}} & \underline{\textbf{41.6}} \\
          & \textbf{\OSRM*} & 4.60  & 26.8  & 60.0  & 307   & 286   & 521   & 218   & 292   & 6.77  & 71.2  & 2.95  & 4.38  & 28028 & 30.7  & 103   & 391 \\
          & \textbf{\CC*} & 19.0  & 32.9  & 334   & 1527  & 1548  & 2121  & 1218  & 1680  & 19.0  & 218   & 11.4  & 11.0  & 4737  & 83.3  & 453   & 2627 \\
          & \textbf{\RK} & 22.1  & 94.7  & 466   & 2466  & 2619  & 3783  & 1765  & 2509  & 38.7  & 848   & 11.2  & 15.5  & 14048 & 108   & 716   & 4175 \\
          & \textbf{\PHAST} &  15.5 & 66.2 & 287   & 1341  & 1421  & 1865  & 1470  & 2053  & 23.5  & 285   & 8.79 & 9.20  & 11361 & 46.0  & 373   & 2388 \\
          & \textbf{\oursysTab-seq} & 29.3  & 89.6  & 383   & 1411  & 1347  & 1823  & 1816  & 2620  & 65.4  & 751   & 18.9  & 26.1  & 7454  & 125   & 545   & 2676 \\
          & \textbf{\OSRM-seq} & 33.1  & 183   & 503   & 3246  & 2839  & 4101  & 3817  & 5273  & 34.4  & 390   & 28.4  & 18.3  & >30000  & 82.4  & 1062  & 12082 \\
          & \textbf{\CC-seq} & 29.2  & 119   & 553   & 2907  & 2604  & 3574  & 3658  & 5002  & 35.1  & 423   & 23.7  & 20.5  & 17268 & 133   & 812   & 6855 \\
    \midrule
    \multicolumn{1}{r}{\multirow{5}[2]{*}{\textbf{Query ($\mu$s)}}} & \textbf{*Ours} & 7.99 & 27.9 & 17.2 & 93.3 & 60.4 & 137 & 376 & 452 & 5.59 & 7.74 & 6.16 & 3.55 & 244 & 1.93 & 10.2 & 33.9 \\
          & \textbf{\OSRM*} & 13.6  & 43.8  & 28.7  & 163   & 97.6  & 223   & 454   & 512   & 7.94  & 9.74  & 6.37  & 3.84  & 325   & 2.20  & 12.5  & 39.0 \\
          & \textbf{\CC*} & 22.1  & 96.4  & 48.1  & 317   & 190   & 424   & 910   & 1010  & 7.60  & 9.07  & 6.73  & 3.33  & 783   & 2.11  & 17.2  & 85.2 \\
              & \textbf{\RK} & \underline{\textbf{6.65}}& \underline{\textbf{24.0}} & \underline{\textbf{13.6}} & \underline{\textbf{79.1}} & \underline{\textbf{47.9}} & \underline{\textbf{102}}& \underline{\textbf{292}} & \underline{\textbf{337}} & \underline{\textbf{4.93}} & \underline{\textbf{6.33}} & \underline{\textbf{4.63}} & \underline{\textbf{2.69}} & \underline{\textbf{176}} & \underline{\textbf{1.52}} & \underline{\textbf{6.25}} & \underline{\textbf{19.0}} \\
          & \textbf{\PHAST} &  12.6    &  39.2   & 26.7  & 138   & 84.4  & 169   & 430   & 482   & 7.73  & 9.23  & 6.98     & 4.09  & 278   & 2.37  & 11.8  & 33.5 \\
          \end{tabular}%
    \caption{\textbf{Comparison of build time (in seconds) and query time (in microseconds) for all tested implementations across all graphs.}
    ``\oursysTab'' $=$ our algorithm.
    ``\OSRM'' $=$ Open Source Routing Machine~\cite{luxen2011real}.
    ``\CC'' $=$ CH-Constructor~\cite{chconstructor2024code}.
    ``\RK'' $=$ RoutingKit~\cite{dibbelt2016customizable}.
    ``\PHAST'' $=$ PHAST~\cite{delling2013phast}.
    Implementations with asterisks (*) are parallel,
    and their sequential running times (with suffices ``-seq'') are also included in the table.
    We report the parallel running time of \CC{} using eight threads,
    as this configuration yields the best performance due to its limited scalability.
    For each graph, the best build time and best query time are bolded and underlined in their respective sections.
}
\label{tab:ch_algo_cmp}%
\end{table*}}{
\begin{table*}[htbp]
  \small
  \centering
  \setlength{\tabcolsep}{3pt}
    \begin{tabular}{cl|c@{\,\,\,}c@{\,\,\,}c@{\,\,\,}c@{\,\,\,}c@{\,\,\,}c|c@{\,\,\,}c@{\,\,\,}c@{\,\,\,}c|c@{\,\,\,}c@{\,\,\,}c@{\,\,\,}c@{\,\,\,}c@{\,\,\,}c}
    \multicolumn{2}{c|}{\multirow{2}[0]{*}{\textbf{Graph}}} & \multicolumn{6}{c|}{\textbf{Road}}            & \multicolumn{4}{c|}{\textbf{Synthetic}} & \multicolumn{6}{c}{\textbf{$k$-NN}} \\
    \multicolumn{2}{c|}{} & \textbf{CA} & \textbf{AO} & \textbf{AF} & \textbf{NA} & \textbf{AS} & \textbf{EU} & \textbf{TRCE} & \textbf{BUB} & \textbf{CHN7} & \textbf{CHN8} & \textbf{HT5} & \textbf{CH2} & \textbf{CH5} & \textbf{GL2} & \textbf{GL5} & \textbf{GL10} \\
    \multicolumn{2}{c|}{\textbf{\# vertices}} & 3.34M & 6.22M & 33.5M & 87.0M & 95.7M & 131M  & 16.0M & 21.2M & 10.0M & 100M  & 929K  & 4.21M & 4.21M & 24.9M & 24.9M & 24.9M \\
    \multicolumn{2}{c|}{\textbf{\# edges}} & 4.20M & 8.24M  & 44.8M & 113M  & 123M  & 169M  & 48.0M & 63.6M & 10.0M & 100M  & 4.64M    & 8.42M & 21.0M & 49.8M & 124M  & 249M \\
    \midrule
    \multicolumn{1}{r}{\multirow{8}[2]{*}{\begin{sideways}\textbf{Build (s)}\end{sideways}}} &  \textbf{\oursysTab*} & \underline{\textbf{1.00}} & \underline{\textbf{2.25}} & \underline{\textbf{7.32}} & \underline{\textbf{23.1}} & \underline{\textbf{23.5}} & \underline{\textbf{32.9}} & \underline{\textbf{25.2}} & \underline{\textbf{34.7}} & \underline{\textbf{1.77}} & \underline{\textbf{14.9}} & \underline{\textbf{0.68}} & \underline{\textbf{0.87}} & \underline{\textbf{115}} & \underline{\textbf{2.95}} & \underline{\textbf{9.78}} & \underline{\textbf{41.6}} \\
          & \textbf{\OSRM*} & 4.60  & 26.8  & 60.0  & 307   & 286   & 521   & 218   & 292   & 6.77  & 71.2  & 2.95  & 4.38  & 28028 & 30.7  & 103   & 391 \\
          & \textbf{\CC*} & 19.0  & 32.9  & 334   & 1527  & 1548  & 2121  & 1218  & 1680  & 19.0  & 218   & 11.4  & 11.0  & 4737  & 83.3  & 453   & 2627 \\
          & \textbf{\RK} & 22.1  & 94.7  & 466   & 2466  & 2619  & 3783  & 1765  & 2509  & 38.7  & 848   & 11.2  & 15.5  & 14048 & 108   & 716   & 4175 \\
          & \textbf{\PHAST} &  15.5 & 66.2 & 287   & 1341  & 1421  & 1865  & 1470  & 2053  & 23.5  & 285   & 8.79 & 9.20  & 11361 & 46.0  & 373   & 2388 \\
          & \textbf{\oursysTab-seq} & 29.3  & 89.6  & 383   & 1411  & 1347  & 1823  & 1816  & 2620  & 65.4  & 751   & 18.9  & 26.1  & 7454  & 125   & 545   & 2676 \\
          & \textbf{\OSRM-seq} & 33.1  & 183   & 503   & 3246  & 2839  & 4101  & 3817  & 5273  & 34.4  & 390   & 28.4  & 18.3  & >30000  & 82.4  & 1062  & 12082 \\
          & \textbf{\CC-seq} & 29.2  & 119   & 553   & 2907  & 2604  & 3574  & 3658  & 5002  & 35.1  & 423   & 23.7  & 20.5  & 17268 & 133   & 812   & 6855 \\
    \midrule
    \multicolumn{1}{r}{\multirow{5}[2]{*}{\begin{sideways}\textbf{Query ($\mu$s)}\end{sideways}}} & \textbf{*Ours} & 7.99 & 27.9 & 17.2 & 93.3 & 60.4 & 137 & 376 & 452 & 5.59 & 7.74 & 6.16 & 3.55 & 244 & 1.93 & 10.2 & 33.9 \\
          & \textbf{\OSRM*} & 13.6  & 43.8  & 28.7  & 163   & 97.6  & 223   & 454   & 512   & 7.94  & 9.74  & 6.37  & 3.84  & 325   & 2.20  & 12.5  & 39.0 \\
          & \textbf{\CC*} & 22.1  & 96.4  & 48.1  & 317   & 190   & 424   & 910   & 1010  & 7.60  & 9.07  & 6.73  & 3.33  & 783   & 2.11  & 17.2  & 85.2 \\
              & \textbf{\RK} & \underline{\textbf{6.65}}& \underline{\textbf{24.0}} & \underline{\textbf{13.6}} & \underline{\textbf{79.1}} & \underline{\textbf{47.9}} & \underline{\textbf{102}}& \underline{\textbf{292}} & \underline{\textbf{337}} & \underline{\textbf{4.93}} & \underline{\textbf{6.33}} & \underline{\textbf{4.63}} & \underline{\textbf{2.69}} & \underline{\textbf{176}} & \underline{\textbf{1.52}} & \underline{\textbf{6.25}} & \underline{\textbf{19.0}} \\
          & \textbf{\PHAST} &  12.6    &  39.2   & 26.7  & 138   & 84.4  & 169   & 430   & 482   & 7.73  & 9.23  & 6.98     & 4.09  & 278   & 2.37  & 11.8  & 33.5 \\
          \end{tabular}%
    \caption{\textbf{Comparison of build time (in seconds) and query time (in microseconds) for all tested implementations across all graphs.}
    ``\oursysTab'' $=$ our algorithm.
    ``\OSRM'' $=$ Open Source Routing Machine~\cite{luxen2011real}.
    ``\CC'' $=$ CH-Constructor~\cite{chconstructor2024code}.
    ``\RK'' $=$ RoutingKit~\cite{dibbelt2016customizable}.
    ``\PHAST'' $=$ PHAST~\cite{delling2013phast}.
    Implementations with asterisks (*) are parallel,
    and their sequential running times (with suffices ``-seq'') are also included in the table.
    We report the parallel running time of \CC{} using eight threads,
    as this configuration yields the best performance due to its limited scalability.
    For each graph, the best build time and best query time are bolded and underlined.
}
\label{tab:ch_algo_cmp}%
\end{table*}}%

\subsection{Experimental Setup}\label{sec:setup}
\myparagraph{Environment.} We run our experiments on a machine machine with four Intel Xeon
Gold 6252 CPUs (96 cores and  192 hyperthreads) and 1.5TB of main memory.
We use \texttt{numactl -i all} for parallel tests to interleave the memory pages across CPUs in a round-robin fashion.

We implement our algorithms in C++ using ParlayLib~\cite{blelloch2020parlaylib} for fork-join parallelism and parallel primitives, which is recently used in many papers on parallel algorithms~\cite{gu2023parallel,shen2022many,BBFGGMS16,blelloch2022parallel}.
ParlayLib is an algorithmic library with parallel building blocks, and also include a scheduler for support fork-join parallelism.
\oursys{} uses some of them such as sorting from ParlayLib, and
uses the ParlayLib scheduler by default, which gives good performance in general.
ParlayLib infrastructure also supports alternative schedulers (e.g., TBB~\cite{TBB} and OpenCilk~\cite{opencilk}).
A comparison of different schedulers for \oursys{} is presented in~\cref{sec:scheduler}.

\myparagraph{Benchmark Dataset.} To evaluate the performance of CH construction algorithms, we use 16 directed graphs.
This dataset includes six road graphs, four synthetic graphs and six $k$-NN graphs.

\begin{itemize}
\item \textbf{Road graphs} are the primary use application domain for CH algorithms. Here we use Central America (CA), Australia Oceania (AO), Europe (EU), North America (NA), Asia (AS), and Africa (AF) from OpenStreetMap~\cite{roadgraph}. The edge weights are natural weights from the source data, which are up to $2^{25}$.
\item \textbf{Synthetic graphs} with relatively low average degrees and high diameters are selected because these characteristics are ideal for testing the performance of CH algorithms.
Here we use hugebubbles-0020 (BUB) and hugetrace-0020 (TRCE) from \cite{sanders2014benchmarking}, which feature 2D adaptively refined triangular meshes.
We assign random weights for BUB and TRCE from 0 to $10^5$.
Chain graphs have $10^7$ (CHN7) and $10^8$ (CHN8) vertices with random weights from 1 to 32.
Whenever a vertex is contracted in the chain, exactly one shortcut is added. Therefore, chain graphs are primarily used to test the cost related to non-scoring steps.
\item \textbf{$\boldsymbol{k}$-NN graphs} are used to evaluate the performance among different average degrees. In $k$-NN graphs, each vertex has $k$ out-going edges pointing to its $k$-nearest neighbors, excluding itself. As $k$ increases, the complexity of the scoring process in CH construction also increases.
We use Humidity and Temperature with $k=5$ (HT5)~\cite{huerta2016online}, Chemical with $k=2,5$ (CH2, CH5)~\cite{fonollosa2015reservoir,wang2021geograph}, and GeoLife with $k=2,5,10$ (GL2, GL5, GL10) \cite{wang2021geograph,geolife}.
Edge weights are randomly assigned from 1 to 32.
\end{itemize}


When we compare the \emph{average} performance across all graphs, we always report the \emph{geometric mean} values across all graphs.

\begin{figure*}
  \centering
    \includegraphics[width=\textwidth]{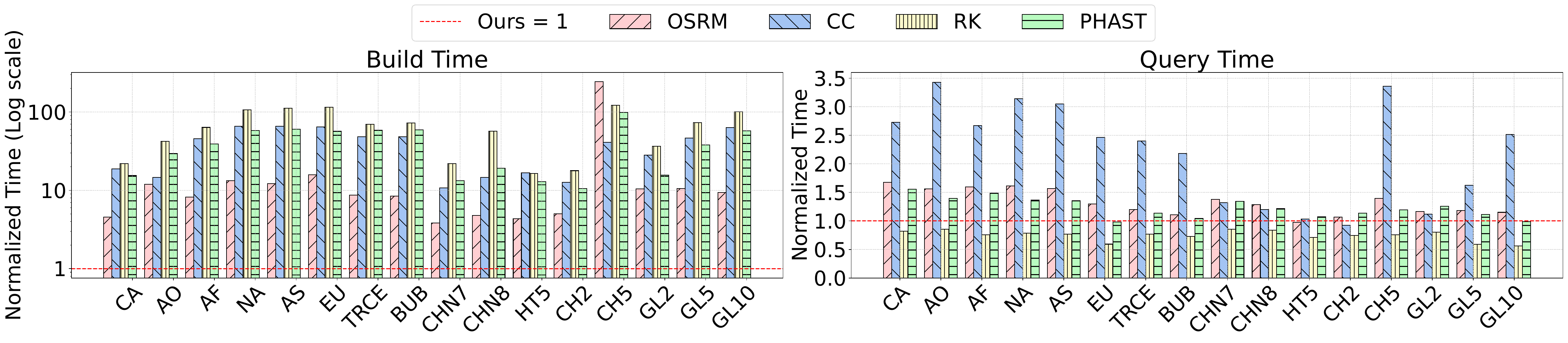}
    \caption{\small \textbf{Build time and query time of all tested baselines, normalized to our algorithm.}
    Lower is better.
    Since all running times are normalized to our algorithm,
    ``\oursysTab{}'' is always equal to one, represented by the horizontal red dotted line.
    }\label{fig:performance_comparison_normalized_graph}
\end{figure*}

\myparagraph{Baseline Competitors.} We call all existing algorithms that we compare to as \defn{baselines}.
We compare \oursys{} to two of the most well-received sequential implementations (\RoutingKit{} and \PHAST) and two state-of-the-art parallel open-source implementations (\chconstructor{} and \OSRM), described as follows.
\begin{itemize}
    \item \textbf{\osrm}~\cite{OSRM,luxen2011real}: A parallel routing engine designed to calculate the fastest routes between locations in the OpenStreetMap~\cite{roadgraph}.
    \item \textbf{\chconstructor~(\chc)}~\cite{chconstructor2024code}: A parallel CH construction implementation. As shown in \cref{fig:parallel_algorithms_scalability_comparison}, \chc{} always achieves the best performance with around 8 threads.
    Hence, the reported construction time for \chc{} is on $8$ threads instead of $192$ threads.
    \item \textbf{\RoutingKit~(\RK)}~\cite{RoutingKit,dibbelt2016customizable}: An efficient sequential C++ library that applies the CH for fast route planning.
    \item \textbf{\PHAST}~\cite{PHAST,delling2013phast}: A widely used algorithm for parallel single-source shortest paths (SSSP) using CH.
    Its implementation includes a sequential process for constructing the CH, and a parallel process for running SSSP.
    In our experiment, we only compare to their construction time, which is sequential.
\end{itemize}

The basic frameworks of both parallel baselines (\chc{} and \osrm) are based on Vetter~\cite{vetter2009parallel}, which supports node scoring and contracting in parallel.
Each time they score a vertex, both algorithms always perform witness searches on all of its neighbors, and they do not record any distance information from these searches.
This means that if scoring two vertices involves performing WPSes on the same vertex, both implementations will perform this search twice,  greatly increasing the total computation.

Since our main focus is on the performance of constructing CH, we obtain the CH generated from all baselines,
and always use the same query algorithm on these CH structures to enable a fair comparison of the CH quality.
We choose to use the query algorithm in \rk{} since we observe that it has the best performance in general.

\subsection{Overall Performance}\label{sec:overall_perform}
\myparagraph{Setup.} 
We conducted experiments to compare performance in terms of construction time and query time with established baselines. Lower time consumption in both indicates better algorithm quality.
\cref{tab:ch_algo_cmp} presents a comparison among \oursys{} and all baselines.
In \cref{fig:performance_comparison_normalized_graph}, we normalize the performance of all baselines to that of \oursys{} (with \oursys{}'s numbers always set to 1).
For construction, we also include the sequential running times of all parallel implementations in \cref{tab:ch_algo_cmp} to reflect the total work.
For queries, for each graph, we randomly select $1000$ pairs of vertices, run these queries sequentially, and report the average time and the average number of vertices processed in the queries.
As mentioned, we apply the same query algorithm to all baselines.

Note that different construction algorithms may generate different CHs on the same input graph.
In particular, a sequential algorithm always chooses the best node to contract in each iteration, while
a parallel algorithm may choose to contract multiple vertices together.
Conceptually, this means that many of the vertices are contracted earlier than they should be.
Therefore, the quality of the CH may be sacrificed due to parallel construction.
To measure the quality of the CHs generated by different algorithms, we use the query time as an indicator.
More precisely, since all algorithms are using the same query algorithm, a lower query time indicates that the CH
structure itself has higher quality.

\hide{
\myparagraph{Memory Usage and Other Measurements.}
\ifconference{In the full version of the paper~\cite{ch-full}, we also report the number of edges in the CH and the number of vertices processed in queries for all baselines.}\iffullversion{We report the number of edges in the CH and the number of vertices processed in queries for all baselines in the appendix.}
The number of edges in CH reflects the space usage for storing the CH. In general, we observe that all
implementations generate similar numbers of edges in the CH.
On almost all graphs, the difference is within 10\%.
The number of vertices processed in queries is generally consistent with the query time.
Therefore, we focus on analyzing the construction and query time here.
}

\myparagraph{Construction Time Evaluation.}\cref{fig:performance_comparison_normalized_graph} and \cref{tab:ch_algo_cmp}
show that our approach achieves significantly better construction performance than all baselines, while maintaining high CH graph quality.
Even our sequential construction time is competitive to the highly-optimized sequential baselines, which is within 1.7$\times$ slower than RK and 2.9$\times$ than PHAST, and can also be faster than them by up to 2.1$\times$.
This indicates that our algorithm incurs very small overhead in work to enable parallelism.
When running in parallel, the construction time of our algorithm is significantly faster than all baselines.
Compared to the best parallel baseline \osrm{}, \oursys{} is 3.8--243$\times$ faster in construction, with an average of 9.9$\times$ faster.
\CC{} suffers from scalability issue in the \step{Reset} step in each round (we will discuss more in \cref{sec:scalability} and \cref{sec:exp:round}).
As a result, it achieves the best running time with 8 threads, which is close to the performance on sequential algorithms.
In general, \oursys{} is 10.7--66.1$\times$ faster than \chc{}, with an average of 31.5$\times$ faster.
As shown in \cref{fig:parallel_algorithms_scalability_comparison},
even only considering the time for scoring and contracting, \chc{} is still not competitive to \osrm{} and is much slower than \oursys{}.
In general, \oursys{} is at least 3.8$\times$ faster than all baselines on all graphs.
On average across all graphs, \oursys{} is 9.9$\times$ faster than \osrm{}, 31.5$\times$ faster than \CC{},
32.4$\times$ faster than (sequential) \phast{}, and 53.8$\times$ than (sequential) \RK{}.

We then analyze the query performance.
As mentioned, the quality for a CH generated by a parallel algorithm is generally expected to be lower than those generated sequentially.
The results indicate that our algorithm still achieves competitive quality to the best sequential implementation.
In general, \RK{} achieves the best query performance in both time and number of vertices visited.
\oursys{} is competitive to \RK, with query times in 1.17--1.79$\times$,
and is better than all other baselines in query time, even including the sequential algorithm \phast{}.
On average, our query time is 1.31$\times$ slower than \rk{}, but is 1.25$\times$ faster than \phast{}, 1.35$\times$ faster than \osrm{}, and 2.07$\times$ faster than \CC{}.
This indicates the CH generated by our parallel algorithm has close quality to a highly-optimized sequential algorithm, and is better than other parallel versions.

\begin{figure}[t]
  \centering
    \includegraphics[width=\columnwidth]{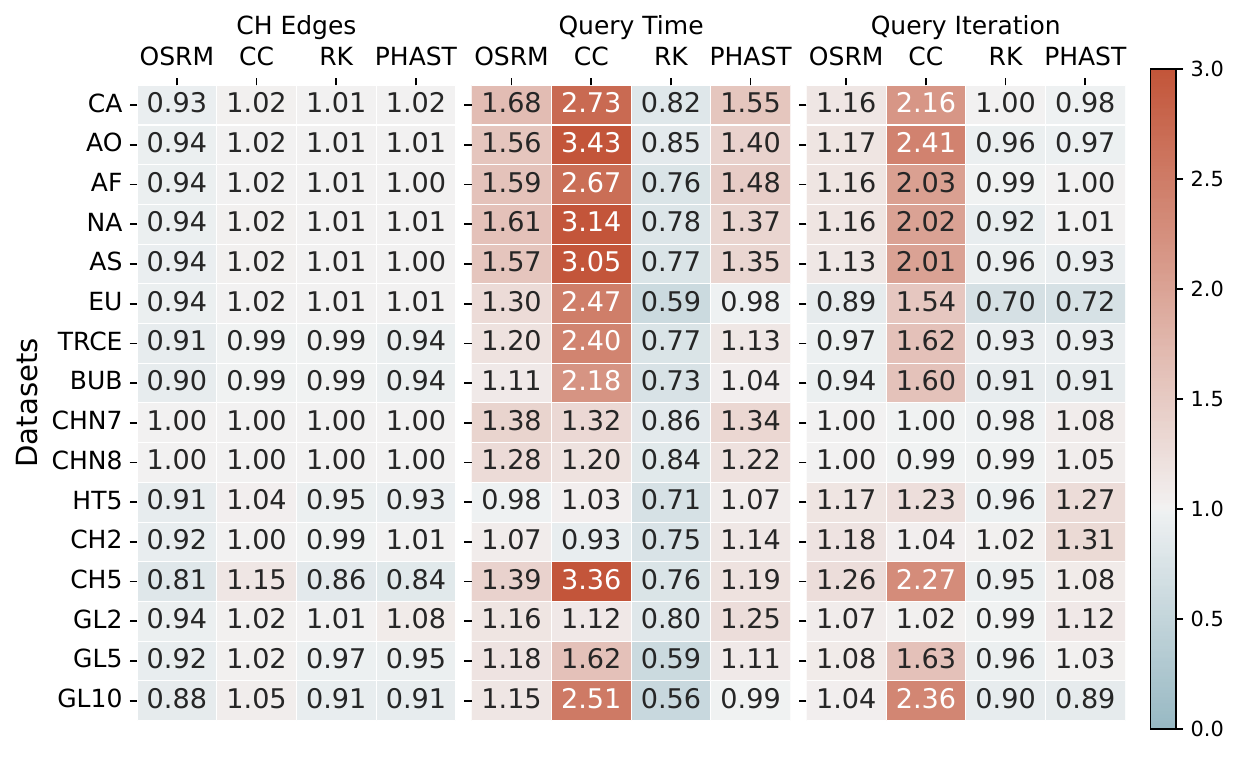}
    \caption{\small \textbf{Heatmap of number of CH edges, query time, and query iteration.}
    All numbers are normalized to ours. Bluer or smaller is better.
    }\label{fig:heatmap_norm}
\end{figure}
\myparagraph{Number of CH Edges, Query Time, and Query Iterations.}\label{sec:ch_edges_query_iteration}
To compare the number of CH edges, query time, and query iterations of our algorithm against the baselines (\OSRM~\cite{luxen2011real}, \CC~\cite{chconstructor2024code}, \RK~\cite{dibbelt2016customizable}, \PHAST~\cite{delling2013phast}),
we present a heatmap in~\cref{fig:heatmap_norm}.
Here ``query iterations'' refers to the average number of vertices visited in an $s$-$t$ query,
which is machine-independent and roughly indicates the query cost.
The numbers in the heatmap are normalized to those of our algorithm.
For CH edges, the differences between the baselines and our algorithm are all within 20\%,
indicating that the space required to store the output CH is similar.
For query time and iterations, \RK{} and \PHAST{}, being sequential and following a stricter contraction order,
perform slightly better than the parallel implementations.
In the worst case, our algorithm is only about twice as slow as the fastest query time. However, since queries finish in microseconds ($10^{-6}$ seconds),
this twofold slowdown is negligible.
Among all parallel implementations, our algorithm achieves the best average query time and iterations.
In summary, our algorithm offers significantly faster construction performance while maintaining competitive output graph size and query time.



\begin{figure}[t]
  \centering
  \includegraphics[width=\columnwidth]{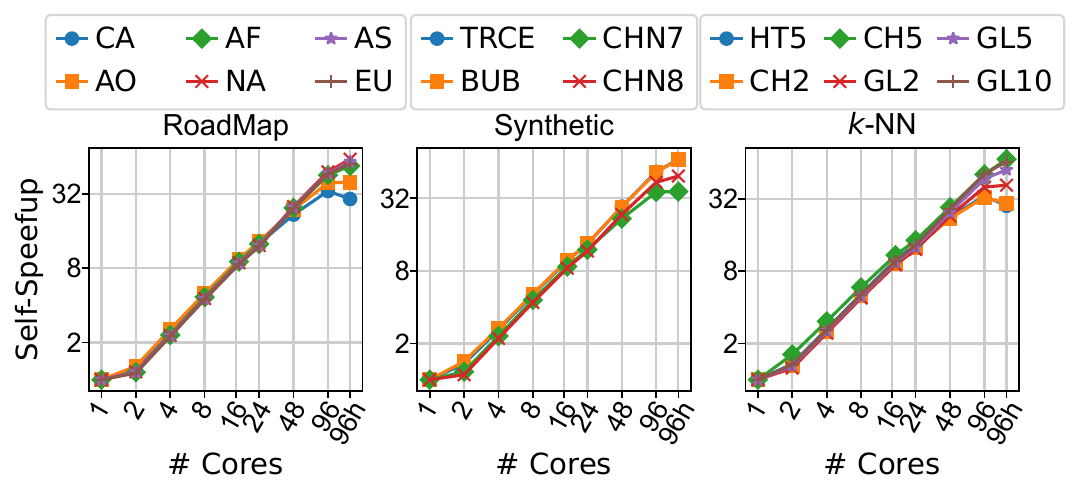}
    \caption{\small \textbf{Self-relative speedups of our algorithm on the three graph categories.} Higher is better.
    \hide{The $y$-axis represents the speedup over the algorithm running sequentially, higher is better. The $x$-axis represents the number of cores. ``96h'' demotes 96 cores with hyperthreading enabled.}
    }\label{fig:scalability}
\end{figure}

\subsection{Scalability}\label{sec:scalability}

\myparagraph{Self-Relative Speedup.}
We first show the self-relative speedup of our algorithms on all the graphs in \cref{fig:scalability}.
We vary the number of processors from $1$ to $96$h ($192$ hyperthreads). Our self-relative speedup is $29.1$--$61.7\times$ on road graphs, $37.9$--$70.8\times$ on synthetic graphs and $27.9$--$67.1\times$ on $k$-NN graphs.
This shows that our algorithm achieves high parallelism on all tested graphs.

\begin{figure}[t]
  \centering
    \includegraphics[width=\columnwidth]{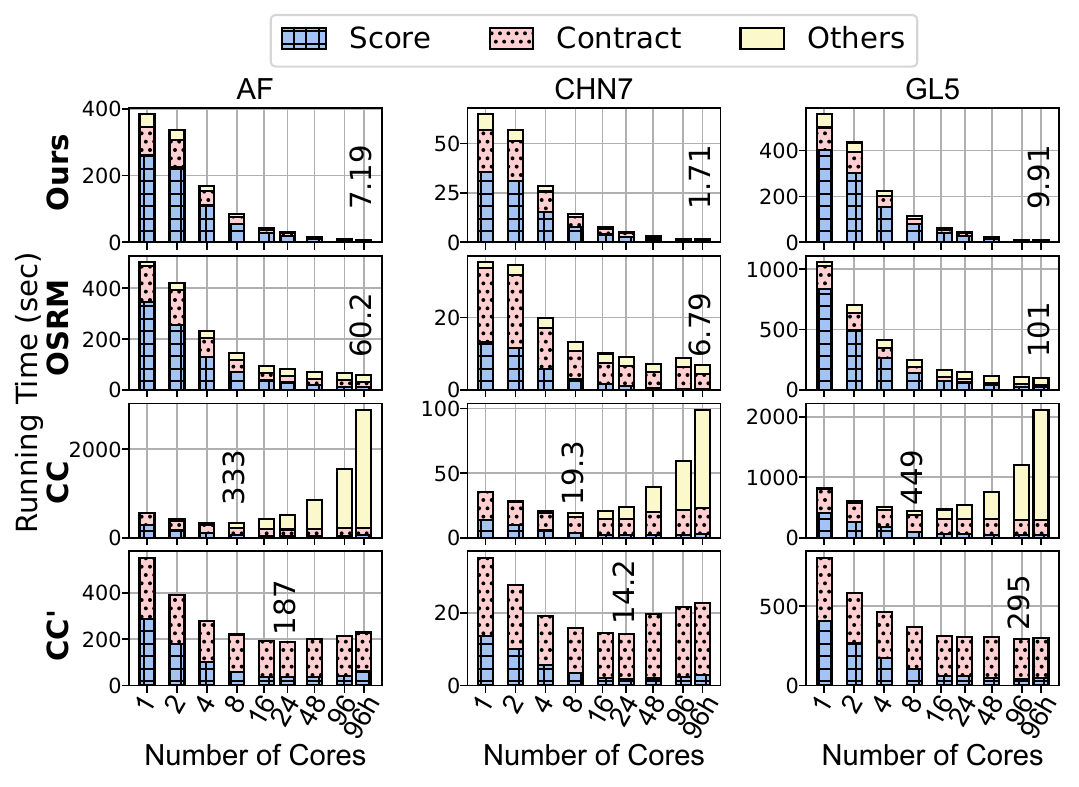}
    \caption{\small \textbf{Running time of \oursys{} (Ours), \osrm{} and \chconstructor{} (\CC) across different core counts.}
    Lower is better.
    The fastest running time across all core counts is given above its bar.
    ``96h'': 96 cores with hyper-threads.
    As the ``Others'' cost in \CC{} dominates in most cases,
    we also report a version of \CC{} that excludes this time, referred to as \CC'.
    }\label{fig:parallel_algorithms_scalability_comparison}
\end{figure}

\myparagraph{Scalability Breakdown.} To validate our claims of improving parallelism in CH preprocessing and to illustrate the limitations of existing state-of-the-art parallel implementations, we test the running times of three parallel algorithms---\oursys{}, \CC, and \OSRM---with cores varying from $1$ to $96$h ($192$ hyperthreads) in \cref{fig:parallel_algorithms_scalability_comparison}.
We select three graphs: AF, CHN7, and GL5, each from a different category.
As mentioned, all parallel implementations roughly follow Vetter's algorithm, so we split the total time into three parts:
\step{Score} (running WPSes and recomputing vertex scores),
\step{Contracting} (finding an independent set, contracting them and adding shortcuts),
and \step{Others}.

\CC{} does not achieve satisfactory scalability, and achieves its best performance with 8 threads.
The major issue lies in a \step{Reset} step that clears arrays at the beginning of each round.
This cost increases drastically with the number of threads. To clarify the breakdown between \step{Score} and \step{Contract},
we also draw a figure for \CC{} without the \step{Others} part, referred to as $\CC'{}$.
The sequential running time for the three algorithms are close.
However, the running time for both \CC{} and \OSRM{} flattens after 16 threads.
Especially for \CC{}, even without considering the resetting time in the \step{Others} part,
its time in \step{Contract} remains the same or even increases when more than 16 threads are used.
This is mainly due to the use of a lock-based structure when adding shortcuts,
since using more threads may result in higher contention and further degrade the performance.
Our solution using parallel hash table effectively avoids this issue, and remains scalable until using all 192 (hyper)threads.

\begin{figure*}[h]
    \begin{subfigure}{\textwidth}
        \centering
        \includegraphics[width=\textwidth]{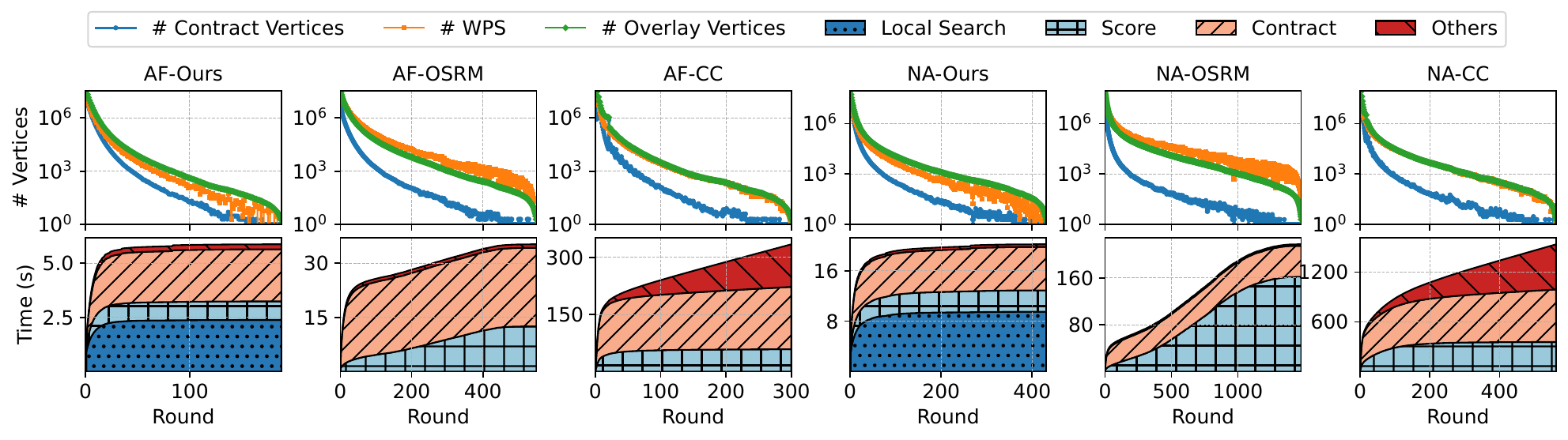}
    \end{subfigure}
    \begin{subfigure}{\textwidth}
        \centering
         \includegraphics[width=\textwidth]{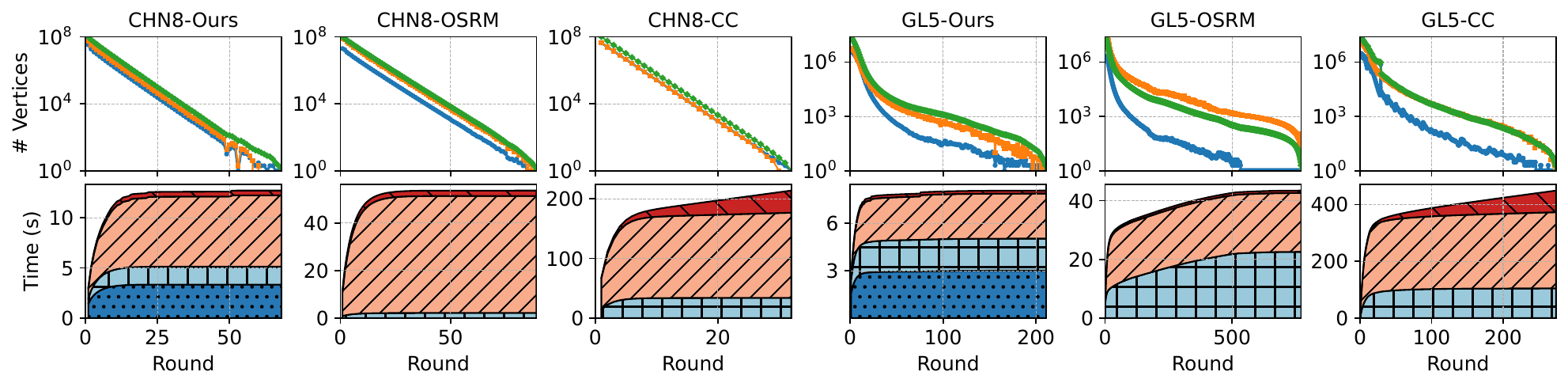}
    \end{subfigure}
    \caption{\textbf{Per-round breakdown of \oursys{} (Ours), \osrm{} and \chconstructor{} (\CC) on four graphs.}
    The x-axis shows the round number.
    The top row shows the number of contracted vertices (blue circles), the number of WPSes (orange squares), and the number of remaining overlay vertices (green diamonds) in each round.
    The bottom row shows the accumulated time for local search (dark blue), scoring (light blue), contracting (coral), and others (red).
    }
    \label{fig:per_round_breakdown}
\end{figure*} 

\OSRM{} has reasonable scalability up to 16 threads, and its running time remains the same when more threads are used.
One possible reason lies in the parallel granularity for computing WPS---\OSRM{} employs parallelism on the level of all vertices in $\AffectedNeighbors$ (vertices that require score recomputing). 
In particular, all vertices in $\AffectedNeighbors$ are parallelized,
but each vertex $v\in \AffectedNeighbors$ itself is processed sequentially, which includes a total of $\nin(v)$ executions of Dijkstra's algorithm.
In later rounds when most vertices have been contracted, there are only a few vertices in $\AffectedNeighbors$,
but their degree may have become large, causing a huge amount of work executed sequentially.
Therefore, its poor performance is due to insufficient parallelism in later rounds.
This issue is also reflected in \cref{fig:per_round_breakdown}, which we discuss in \cref{sec:exp:round}.
In this case, the WPSes are close to a sequential execution, and do not benefit from having more threads.
Our solution that batches all WPS sources and parallelizes them as a whole effectively avoids this issue, and can efficiently utilize more threads.

On the three tested graphs, \oursys{} shows high scalability up to 96h, and always has better performance with more threads.
The self-relative speedup is $53.3\times$ on AF, $37.9\times$ on CHN7, and $56.3\times$ on GL5.
All techniques in \oursys{} are carefully optimized for high parallelism,
such as using lock-free data structures and batching WPSes to run in parallel.
Therefore, with a reasonable sequential cost, the good scalability guarantees low parallel running time.

\subsection{In-Depth Study for Parallel Algorithms}\label{sec:exp:round}

\hide{
\begin{figure*}
    \centering
      \includegraphics[width=0.8\textwidth]{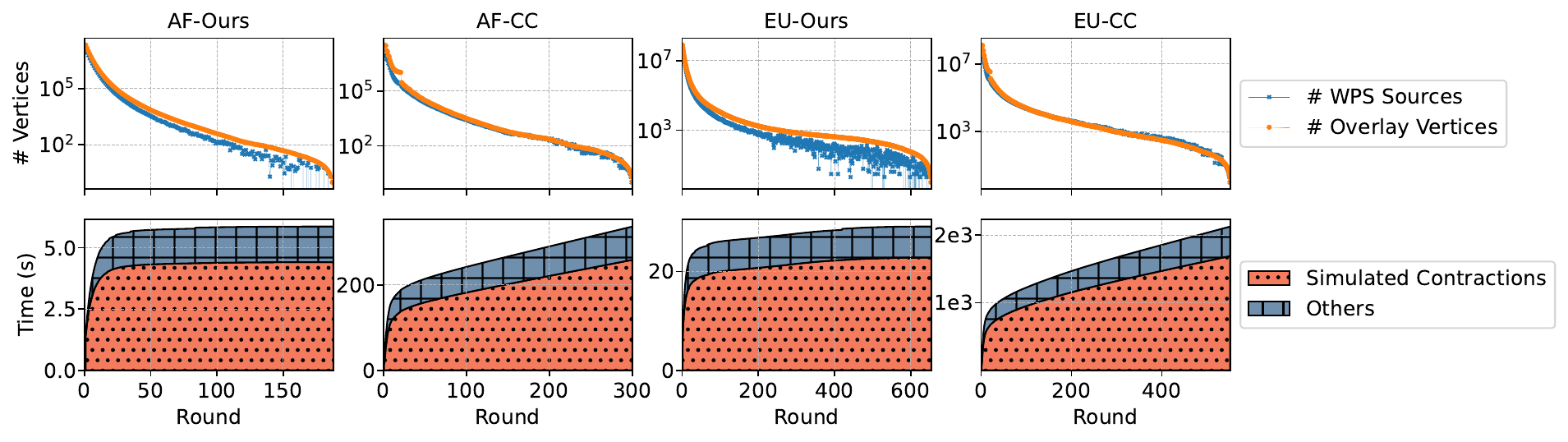}
      \caption{\small \textbf{Number of updated vertices and overlay vertices of each round comparison.\zijin{add comparison of k-NN graph?}}
      }\label{fig:overlay_detail_per_round}
\end{figure*}

\begin{figure*}
	\begin{minipage}{\textwidth}
      \includegraphics[width=\textwidth]{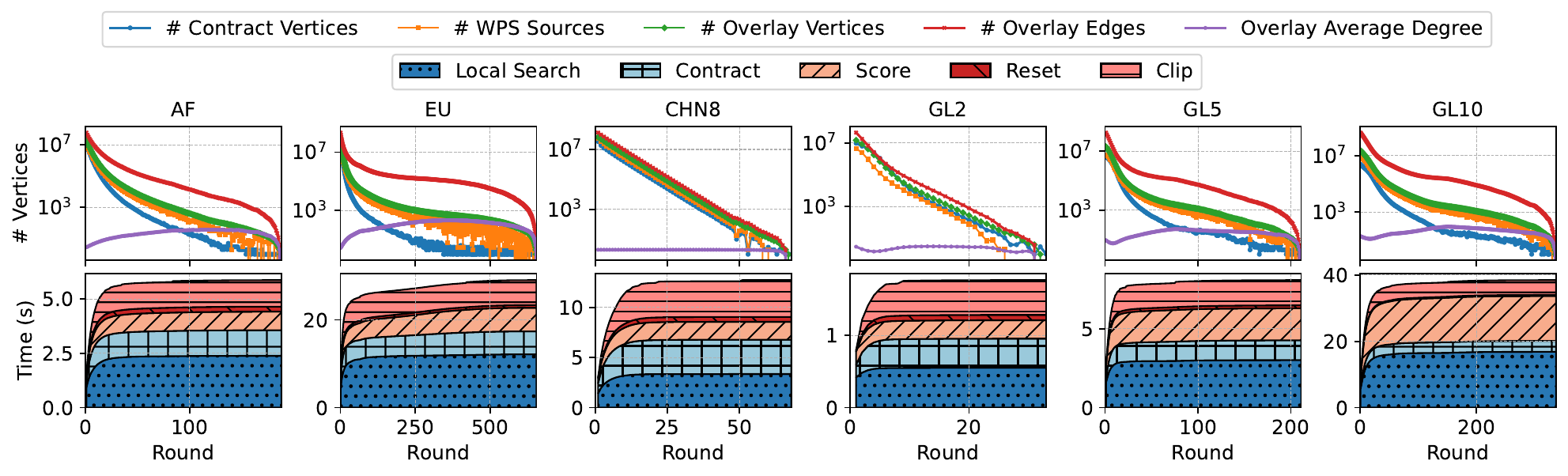}
      \caption{\small \textbf{Ours breakdown of each round.}
      }\label{fig:ours_round_breakdown}
	\end{minipage}
\end{figure*}
}
Recall that parallel CH construction algorithms select an independent set in each round, contract them all in parallel, and repeat until the graph becomes empty.
Therefore, the performance and statistics of each round may help to better understand the process of parallel algorithms.
In this section, we present an in-depth study for the performance breakdown in each round, for all parallel implementations \oursys{}, \CC{} and \OSRM{}.
We present the results on four graphs in \cref{fig:per_round_breakdown}, which includes the number of vertices processed and accumulated running time up to each round.

\myparagraph{Number of Vertices Processed.}
In \cref{fig:per_round_breakdown}, we present the accumulated numbers of vertices left in the overlay graph ($|\overlayV|$), the number of vertices contracted ($\FeasibleCandidates$),
as well as the number of WPSes (i.e., the number of executions of Dijkstra's algorithm).
We first note that \osrm{} incurs much more rounds than \oursys{} and \CC{}---up to 4$\times$ more rounds than \oursys{}.
The larger number of rounds for \osrm{} is caused by how the independent set is selected. \osrm{} selects a vertex when it has the minimum score in its 2-hop neighborhood, while \oursys{} uses 1-hop neighborhood. The goal of using 2-hop neighborhood is to avoid using complicated concurrent data structure to deal with shortcut insertion, but it results in a much slower contraction process than \oursys{}.
On the contrary, \oursys{} uses the efficient data structure introduced in \cref{sec:ds}, which allows contracting all vertices with minimum score in their 1-hop neighborhood, leading to fewer rounds.

For all three algorithms, the size of the overlay graph decreases quickly.
On CHN8, the trend of $|\overlayV|$ decreases linearly on the log scale.
Indeed on a chain, since almost all vertices have the same edge difference,
the contraction is mostly determined by the random priority. In this case, a constant fraction of vertices can be selected in expectation~\cite{shun2015sequential}.
On other graphs, the size of the overlay graph decreases rapidly in the first few rounds, and then slows down.
In general, for all implementations, the overlay graph size decreases by more than 99\% in the first 10--20 rounds.
This implies great potential of parallelism in the construction of CH.

We then compare the number of WPSes executed for each algorithm.
As mentioned, one of our efforts is to reduce the number of WPSes performed to optimize the overall performance.
\oursys{} incurs significantly fewer Witness Path Searches (WPSes) compared to \CC{} and \osrm.
In each round, especially in later rounds, \oursys{} requires far fewer WPSes than the total number of vertices in the overlay graph.
This reduction in the number of WPS demonstrates the efficiency of our approach in limiting unnecessary searches.

\myparagraph{Performance Breakdown.}
We now analyze the accumulated running time across rounds.
In the first $10\%$ of the rounds, the running time for both algorithms grows quickly.
As mentioned, more than 99\% of the vertices may be contracted just in the first 10--20 rounds.
As the algorithm proceeds, the overlay graph becomes much smaller.
On the four graphs, all algorithms reach a small overlay graph with $10^5$ within 30 rounds,
after which very little computation is required.
Indeed, our algorithm spends the majority of its time in the first several rounds.
As the size of the overlay graph decreases, the running time for the last 90\% of rounds is small.
Indeed, all the steps become cheap after a few rounds.
This indicates the effectiveness of our solution to maintain the overlay graph---when the overlay graph becomes small, the maintenance cost also shrinks proportionally.
This benefit comes from our technique of maintaining the overlay graph in a lazy manner.
In later rounds where only a few shortcuts are generated in each round, \oursys{} delays the process of combining them into the CSR until it collects a sufficient number of shortcuts.
Thus, after $10$--$20$ rounds, there are likely only one or two global merges incurred, leading to \oursys{}'s high overall performance.

In contrast, the running time for \osrm{} grows steadily across all rounds.
This indicates that even processing a small overlay graph, \osrm{} may spend a noticeable amount of time.
As we discussed in \cref{sec:scalability}, this is likely due to insufficient parallelism in later rounds.
For \CC{}, as mentioned, the use of inefficient data structures to maintain shortcuts dramatically increases its running time in the \step{Contract} step,
leading to overall unsatisfactory performance.

In summary, our design allows for contracting a large fraction of vertices in each round, leading to a very fast contraction process.
Our new design for the \step{LocalSearch} step also reduces the number for WPSes performed.
Our data structure also avoids a high cost to maintain a small overlay graph.
As a result, \oursys{} outperforms both parallel baselines due to new designs in both algorithm.

\begin{figure}
    \centering
      \includegraphics[width=\columnwidth]{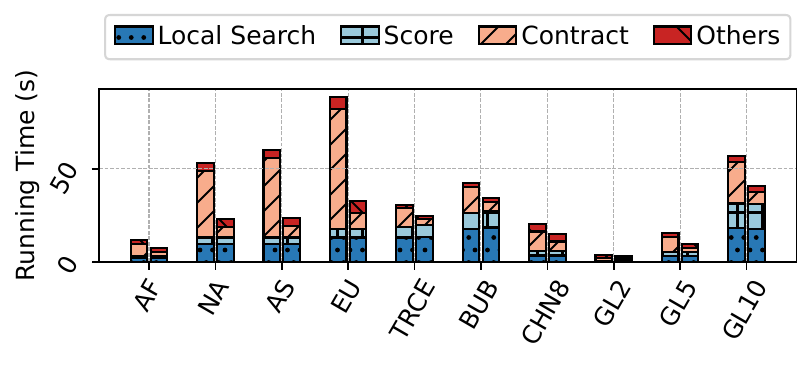}
      \caption{\small \textbf{Running time with and without the Lazy Combination Optimization.}
      For each graph, the left bar corresponds to the cost breakdown without lazy combination, while the right bar represent the breakdown with the optimization.
      The cost of combination is counted in the \step{Contract} step.
      }\label{fig:breakdown_clip}
\end{figure}

\aptLtoX[graphic=no,type=html]{\begin{table*}
  \centering
  \setlength{\tabcolsep}{4pt}
    \begin{tabular}{cl|c@{\,\,}c@{\,\,}c@{\,\,}c@{\,\,}c@{\,\,}c|c@{\,\,}c@{\,\,}c@{\,\,}c|c@{\,\,}c@{\,\,}c@{\,\,}c@{\,\,}c@{\,\,}c}
      \multicolumn{2}{c|}{\multirow{2}[1]{*}{\textbf{Graph}}} & \multicolumn{6}{c|}{\textbf{Road}}            & \multicolumn{4}{c|}{\textbf{Synthetic}} & \multicolumn{6}{c}{\textbf{$k$-NN}} \\
      \multicolumn{2}{c|}{} & \textbf{CA} & \textbf{AO} & \textbf{AF} & \textbf{NA} & \textbf{AS} & \textbf{EU} & \textbf{TRCE} & \textbf{BUB} & \textbf{CHN7} & \textbf{CHN8} & \textbf{HT5} & \textbf{CH2} & \textbf{CH5} & \textbf{GL2} & \textbf{GL5} & \textbf{GL10} \\
      \midrule
      \multicolumn{1}{c}{\multirow{3}[2]{*}{\begin{sideways}\textbf{Build}\end{sideways}}} & \textbf{ParlayLib~\cite{blelloch2020parlaylib}} & 1.00  & 2.25  & \underline{\textbf{7.32}}  & \underline{\textbf{23.1}}  & \underline{\textbf{23.5}}  & 32.9  & 25.2  & 34.7  & \underline{\textbf{1.77}}  & \underline{\textbf{14.9}}  & 0.68  & 0.87  & 115   & \underline{\textbf{2.95}}  & 9.78  & 41.6 \\
            & \textbf{OpenCilk~\cite{opencilk}} & \underline{\textbf{0.97}}  & \underline{\textbf{2.21}}  & 7.37  & 23.1  & 23.7  & \underline{\textbf{32.5}}  & \underline{\textbf{24.9}}  & \underline{\textbf{32.4}}  & 1.82  & 15.4  & 0.67  & \underline{\textbf{0.86}}  & \underline{\textbf{111}}   & 2.96  & \underline{\textbf{9.71}}  & \underline{\textbf{41.2}} \\
            & \textbf{TBB~\cite{TBB}} & 1.03  & 2.33  & 7.66  & 23.5  & 24.0  & 32.8  & 25.1  & 34.3  & 1.93  & 15.9  & \underline{\textbf{0.50}}  & 0.96  & 114   & 3.30  & 10.3  & 42.3 \\
      \midrule
      \multicolumn{1}{c}{\multirow{3}[2]{*}{\begin{sideways}\textbf{Query}\end{sideways}}} & \textbf{ParlayLib} & 7.99  & 27.9  & 17.2  & 93.3  & 60.4  & 137   & \underline{\textbf{376}}   & 452   & 5.59  & \underline{\textbf{7.74}}  & \underline{\textbf{6.16}}  & 3.55  & 244   & 1.93  & \underline{\textbf{10.2}}  & 33.9 \\
            & \textbf{OpenCilk} & 7.91  & \underline{\textbf{27.8}}  & \underline{\textbf{17.1}}  & \underline{\textbf{93.3}}  & 60.7  & 141   & 379   & 454   & \underline{\textbf{5.56}}  & 7.78  & 6.24  & \underline{\textbf{3.53}}  & \underline{\textbf{239}}   & 1.93  & 10.4  & 32.8 \\
            & \textbf{TBB} & \underline{\textbf{7.90}}  & 27.9  & 17.3  & 94.4  & \underline{\textbf{57.6}}  & \underline{\textbf{136}}   & 377 & \underline{\textbf{452}}   & 5.64  & 7.76  & 6.27  & 3.54  & 242   & \underline{\textbf{1.92}}  & 10.3  & \underline{\textbf{32.5}} \\
      \end{tabular}%
  \caption{\textbf{Comparison of build time (in seconds) and query time (in microseconds) for all tested schedulers across all graphs.} Smaller is better. For each graph, the best build time and best query time are bolded and underlined in their respective sections.}
  \label{tab:scheduler}%
\end{table*}}{
\begin{table*}[htbp]
  \centering
  \setlength{\tabcolsep}{4pt}
    \begin{tabular}{cl|c@{\,\,}c@{\,\,}c@{\,\,}c@{\,\,}c@{\,\,}c|c@{\,\,}c@{\,\,}c@{\,\,}c|c@{\,\,}c@{\,\,}c@{\,\,}c@{\,\,}c@{\,\,}c}
    \multicolumn{2}{c|}{\multirow{2}[1]{*}{\textbf{Graph}}} & \multicolumn{6}{c|}{\textbf{Road}}            & \multicolumn{4}{c|}{\textbf{Synthetic}} & \multicolumn{6}{c}{\textbf{$k$-NN}} \\
    \multicolumn{2}{c|}{} & \textbf{CA} & \textbf{AO} & \textbf{AF} & \textbf{NA} & \textbf{AS} & \textbf{EU} & \textbf{TRCE} & \textbf{BUB} & \textbf{CHN7} & \textbf{CHN8} & \textbf{HT5} & \textbf{CH2} & \textbf{CH5} & \textbf{GL2} & \textbf{GL5} & \textbf{GL10} \\
    \midrule
    \multicolumn{1}{c}{\multirow{3}[2]{*}{\begin{sideways}\textbf{Build}\end{sideways}}} & \textbf{ParlayLib~\cite{blelloch2020parlaylib}} & 1.00  & 2.25  & \underline{\textbf{7.32}}  & \underline{\textbf{23.1}}  & \underline{\textbf{23.5}}  & 32.9  & 25.2  & 34.7  & \underline{\textbf{1.77}}  & \underline{\textbf{14.9}}  & 0.68  & 0.87  & 115   & \underline{\textbf{2.95}}  & 9.78  & 41.6 \\
          & \textbf{OpenCilk~\cite{opencilk}} & \underline{\textbf{0.97}}  & \underline{\textbf{2.21}}  & 7.37  & 23.1  & 23.7  & \underline{\textbf{32.5}}  & \underline{\textbf{24.9}}  & \underline{\textbf{32.4}}  & 1.82  & 15.4  & 0.67  & \underline{\textbf{0.86}}  & \underline{\textbf{111}}   & 2.96  & \underline{\textbf{9.71}}  & \underline{\textbf{41.2}} \\
          & \textbf{TBB~\cite{TBB}} & 1.03  & 2.33  & 7.66  & 23.5  & 24.0  & 32.8  & 25.1  & 34.3  & 1.93  & 15.9  & \underline{\textbf{0.50}}  & 0.96  & 114   & 3.30  & 10.3  & 42.3 \\
    \midrule
    \multicolumn{1}{c}{\multirow{3}[2]{*}{\begin{sideways}\textbf{Query}\end{sideways}}} & \textbf{ParlayLib} & 7.99  & 27.9  & 17.2  & 93.3  & 60.4  & 137   & \underline{\textbf{376}}   & 452   & 5.59  & \underline{\textbf{7.74}}  & \underline{\textbf{6.16}}  & 3.55  & 244   & 1.93  & \underline{\textbf{10.2}}  & 33.9 \\
          & \textbf{OpenCilk} & 7.91  & \underline{\textbf{27.8}}  & \underline{\textbf{17.1}}  & \underline{\textbf{93.3}}  & 60.7  & 141   & 379   & 454   & \underline{\textbf{5.56}}  & 7.78  & 6.24  & \underline{\textbf{3.53}}  & \underline{\textbf{239}}   & 1.93  & 10.4  & 32.8 \\
          & \textbf{TBB} & \underline{\textbf{7.90}}  & 27.9  & 17.3  & 94.4  & \underline{\textbf{57.6}}  & \underline{\textbf{136}}   & 377 & \underline{\textbf{452}}   & 5.64  & 7.76  & 6.27  & 3.54  & 242   & \underline{\textbf{1.92}}  & 10.3  & \underline{\textbf{32.5}} \\
    \end{tabular}%
  \caption{\textbf{Comparison of build time (in seconds) and query time (in microseconds) for all tested schedulers across all graphs.} Smaller is better. For each graph, the best build time and best query time are bolded and underlined.}
  \label{tab:scheduler}%
\end{table*}}%

\myparagraph{Evaluating the Lazy Combination of Shortcut Edges.}
As mentioned, one of our optimizations is to combine the shortcuts $\ShortcutEdges$ lazily with the edges of the overlay graph $\overlayE$.
This is to reduce and amortize the cost of updating the CSR for $\overlayE$.
\cref{fig:breakdown_clip} shows the running time with and without this optimization.
For each graph, the left bar represents the breakdown that combines in each round (without lazy combination), while the right bar represents that with the optimization.
The cost is counted in the \step{Contract} step.
By combining the shortcuts lazily instead of doing it every round, the cost of the \step{Contract} step is reduced from 39.9\% to 19.8\% of the overall time on average.
Across all graphs, this optimization improves the performance by 1.1--2.7$\times$ for the total running time.

\myparagraph{Impact of Different Schedulers.}\label{sec:scheduler}
To evaluate the impact of different schedulers to the performance of \oursys{}, we conduct an experiment comparing ParlayLib~\cite{blelloch2020parlaylib}, OpenCilk~\cite{opencilk}, and TBB~\cite{TBB}.
Our algorithmic framework remains the same across all tests, ensuring that any observed performance differences stem primarily from the underlying scheduler rather than the core algorithm itself.
Across all schedulers, the average build time is within $10\%$ of the fastest result, and the average query time is within $2\%$.
Hence, the choice of scheduler has a negligible effect on the performance of \oursys{}.
We present detailed running times in~\cref{tab:scheduler}.

\section{Related Work}\label{sec:related}

Computing the shortest paths on a graph is one of the most well-studied problems in computer science.
We refer the audience to the excellent surveys on this topic, including those by Bast et al.~\cite{bast2016route}, Madkoure et al.~\cite{madkour2017survey}, and Sommer~\cite{sommer2014shortest}.
These surveys all discuss contraction hierarchies (CH) in detail.
Meanwhile, we note that the main applications of the CH are on sparse networks such as roadmaps (transition networks).
We acknowledge that many interesting algorithms have been proposed for relevant but different problems, such as on social networks, queries among all pairs or between a batch of sources and destinations, on dynamic graphs, and time-dependant queries.
A brief list of recent work on computing shortest paths includes~\cite{zhang2022shortest,zhang2021efficient,qiu2022efficient,li2020continuously,yu2020distributed,li2020fast,zhang2020hub,huang2021learning,gong2024querying}.

The idea of CH was proposed by Geisberger et al.~\cite{geisberger2008contraction}, based on simplifying highway hierarchies~\cite{sanders2006engineering, knopp2007computing} and highway node routing~\cite{schultes2007dynamic}.
CH has achieved notable success in practical applications, and fostered numerous later studies on relevant problems.
Examples include the time-dependent versions~\cite{batz2010time,batz2013minimum}, parallel and distributed versions~\cite{kieritz2010distributed,vetter2009parallel}, on dynamic graphs~\cite{ouyang2020efficient}, and more algorithmic optimizations~\cite{geisberger2012exact,delling2017customizable,hespe2019more}.
CH is also used to parallelize SSSP queries, such as in \PHAST{}~\cite{PHAST,delling2013phast}.
Another stream of research is to derive theoretical guarantees for CH~\cite{dibbelt2016customizable,blum2021sublinear,funke2015provable,abraham2010highway,hamann2018graph,columbus2009complexity}.
These analyses are mostly parameterized, based on some graph invariants such as tree depths, tree widths, and diameters.

There have been many studies on parallelizing CH construction and relevant techniques.
Vetter's work~\cite{vetter2009parallel} is the earliest and inspired many of the later studies~\cite{OSRM,luxen2011real,chen2024parallel,karimi2019gpu,karimi2020fast,kieritz2010distributed,chconstructor2024code}.
We reviewed Vetter's approach in \cref{sec:exist-para-ch}.
\OSRM{} by Luxen and Vetter~\cite{OSRM,luxen2011real} implements Vetter's approach, and it is considered the SOTA parallel CH construction.
\chconstructor~\cite{chconstructor2024code} is another open-source software for parallel CH construction.
We compared to \OSRM{} and \chconstructor{} since they have open-source code available.
The others focus on different settings, such as distributed~\cite{kieritz2010distributed}, GPU~\cite{karimi2019gpu,karimi2020fast}, and on edge contraction~\cite{chen2024parallel}. Among them, an existing GPU algorithm~\cite{karimi2020fast} also prune unnecessary shortcuts.
\oursys{} has the same motivation for pruning, but uses a different methodology by leveraging memoization to further reduce the cost.
We believe that some of our algorithmic techniques, such as LocalSearch with memoization, are applicable to other settings such as GPUs. However, we acknowledge that part of the implementation, such as the parallel data structures, will require careful redesign, which we leave as future work.


We note that there are other shortest-path algorithms that can provide different construction-query trade-offs or are designed for other graph types.
Some of these algorithms include transit node routing~\cite{bast2007fast}, hub labeling~\cite{abraham2011hub,abraham2012hierarchical}, pruned landmark labeling~\cite{akiba2013fast}, highway labeling~\cite{akiba2014fast}, and ALT~\cite{goldberg2005computing}.
Indeed, many of these approaches~\cite{arz2013transit,bast2006ultrafast,bast2007fast,abraham2011hub,abraham2012hierarchical,delling2013hub,bauer2010combining,delling2013phast} use CH as a subroutine.
We believe that the faster CH construction presented in this paper can improve such trade-offs, and we leave it as future work.

\section{Conclusions and Future Work}
In this paper, we propose \oursys{} (Scalable Parallelization of Contraction Hierarchies), 
a parallel algorithm for constructing contraction hierarchies. 
Our key insights include algorithm redesign to introduce the \step{LocalSearch} step, 
which allows for batching, memoization and pruning, 
as well as leveraging parallel data structures. 
In this way, \oursys{} effectively reduces the total work and enhances parallelism. 
Across 16 graphs of varying sizes and average degrees (including road networks, synthetic graphs, and $k$-NN graphs), 
\oursys{} consistently outperforms four other SOTA sequential and parallel baselines by 3.83--243$\times$, 
while maintaining competitive query performance.
On a 96-core machine, \oursys{} delivers self-relative speedups of 27.9--70.8$\times$. 
We conduct in-depth experiments to analyze the improvements of our techniques.

An open question remains as to whether parallel CHs can 
handle graphs with high degrees (e.g., social networks). 
Some interesting future directions include applying our new CH construction algorithm to other distance queries, such as single-source shortest paths,
all-pairs shortest paths, or distance oracles.

\begin{acks}
This work is supported by NSF grants CCF-2103483, IIS-2227669, and TI-2346223, NSF CAREER Awards CCF-2238358 and CCF-2339310, the UCR Regents Faculty Development Award, and the Google Research Scholar Program.
We thank the anonymous reviewers for their useful comments.
\end{acks}
\bibliographystyle{ACM-Reference-Format}
\balance
\bibliography{bib/strings, bib/main}

\iffullversion{
\appendix




}

\end{document}
\endinput